\documentclass[useAMS,usenatbib,usegraphicx,referee]{mn2e}

\begin{document}

\title[P-Cygni profiles from outflows near compact objects.]
{Gravitationally distorted P-Cygni profiles from outflows near compact objects.}
\author[A. V. Dorodnitsyn ]{A. V. Dorodnitsyn$^{1,2}$\thanks{E-mail:
dora@milkyway.gsfc.nasa.gov}\\
$^{1}$Laboratory for High Energy Astrophysics, NASA Goddard Space Flight Center, Code 662, Greenbelt, MD, 20771, USA\\
$^{2}$Space Research Institute, Profsoyuznaya st., 84/32, 117997, Moscow, Russia}

\date{}

\pagerange{\pageref{firstpage}--\pageref{lastpage}} \pubyear{2002}
\maketitle

\label{firstpage}
\begin{abstract}

We consider resonant
absorption in a spectral line in the
outflowing plasma within several tens of Schwarzschild radii from
a compact object. 
We take into account both Doppler and gravitational shifting effects 
and re-formulate the
theory of P-Cygni profiles in these new circumstances.
It is found that a spectral  
line may have multiple absorption and emission components
depending on how far the region of interaction is from the compact object and what is the distribution of 
velocity and opacity. Profiles of spectral lines produced near a neutron star or a black hole 
can be strongly distorted by Doppler blue-, or red-shifting, and gravitational red-shifting.
These
profiles may have both red- and blue-shifted absorption troughs.
The result should be contrasted with classical P-Cygni profiles which consist of red-shifted emission and
blue-shifted absorption features. 

We suggest this property
of line profiles to have complicated narrow absorption and emission components in the presence of strong gravity may
help to study spectroscopically
the innermost parts of an outflow. 
\end{abstract}

\begin{keywords}
line formation -- radiative transfer --
galaxies: active -- radiation mechanisms: general -- stars: mass loss -- stars: winds, outflows
\end{keywords}

\section{Introduction}

It is widely accepted that radiation produced in the vicinity of
compact objects can bear an imprint of the strong
gravitational field, of either a stellar mass black hole (BH) or a
neutron star (NS) in Galactic X-ray binary systems, or of a
super-massive BH in case of active galactic nuclei
(AGN). 
Thus the problem of direct detection of line emission from
regions close to compact objects is of great
importance. 
One of the most important piece of evidence for the presence of an accretion disk, and also
the most extensively theoretically investigated subject in this field, is a broadened, red-shifted spectral feature attributed to a
fluorescent ${\rm K}\alpha$ emission line observed in the X-ray spectra of some Seyfert 1 galaxies. It
is believed to be formed due to the reprocessing of X-Rays emitted by
corona in the inner parts of a relatively cold accretion disk (see e.g. ~\cite{Fabian2000}). If this is the case, then the line forming region translates to a broad region in velocity and frequency  space making a line profile
severely skewed by relativistic effects of the accreting plasma.
This strong distortion of the spectral line
makes detailed diagnostics of 
physical conditions in the plasma using this line difficult. 

The grating spectrographs on the X-ray telescopes {\it Chandra} and {\it XMM-Newton} provide unprecedented 
spectral resolution up to $\sim10\, {\rm keV}$.
Thus, by means of {\it Chandra} High Energy Grating observations, a narrow "core" of the Fe ${\rm K}\alpha$ line was detected in many type I AGNs (\cite{Yaqoob2003}).
In order to explain this spectral feature in these sources it is necessary to invoke both reflection from an accretion disk and also emission from distant 
regions (such as an obscuring torus). 

Observations of several quasars and Seyfert galaxies reveal
narrow X-ray absorption lines indicating matter outflowing, falling onto, and orbiting 
near the BH. These features 
are interpreted as Fe absorption
lines, blue-, and red-shifted relative to the rest-frame frequency. 

In some cases, observed spectral features are
interpreted as being gravitationally red-shifted absorption lines.
Tentative evidence has been found for an
absorption feature which presumably originates from resonant
scattering of iron within an accretion flow in the strong
gravitational field of the BH in the Seyfert 1
galaxy NGC 3516 (~\cite{Nandra99}).
In the paper by \citet{Yaqoob} it
was suggested that the detection of the narrow red-shifted
absorption line superimposed on the red wing of a broad Fe ${\rm
K\alpha}$ line in the quasar E1821+643 is due to
resonance absorption by  Fe {XXV} or Fe XXVI, gravitationally red-shifted from region within 
10-20 $r_g$ from the BH. ~\citet{Reeves} interpreted an
absorption features from the quasar PG 1211+143 as being produced
by ionized iron and either being Doppler red-shifted due to the plasma
falling onto the black hole, or due to gravitational red-shifting, or
both. ~\citet{Matt}  also reported on the possible detection of a
transient absorption line from the ionized Fe from the quasar
Q0056-363. If the observed red-shift is only gravitational
then it has been argued  the absorbing gas lies within $\sim 5\,r_g$ from the BH. In the paper by \cite{Dadina05} the red- and blue-shifted absorption lines are found in the X-ray spectrum of the Seyfert 1 galaxy Mrk 509. The data suggest these features are formed in a flow moving with velocity, $v\sim 0.15-0.2\,c$, or/and being gravitationally red-shifted.

Much smaller astrophysical objects which may potentially demonstrate
gravitationally red-shifted lines are neutron stars.
It is widely believed that X-ray bursts are produced as neutron stars are accreting plasma from their massive companions in close binary systems.
It was also suggested that measurement of the gravitationally red-shifted absorption lines during
an X-ray burst can yield the mass-to-radius relation for the neutron star and thus provides the required
constraint to the equation of state of the NS interior.
\citet{Cottam} reported on the observation of several absorption features in the X-ray burst spectra of
the low-mass X-ray binary EXO0748-676.
Prominent absorption features were attributed to the red-shifted lines of
Fe {XXVI}, Fe {XXV}, and O {VIII}. It was speculated that both gravitational red-shift and possibly a slow outflow are responsible for the observed features.

From the above examples one may conclude that there exists the possibility to observe gravitationally red-shifted narrow spectral features from diverse classes of objects.

With an exception of a purely photospheric line, it is natural to assume that plasma which produces line emission is moving with respect to the source of radiation (photosphere or accretion disk).

The line emitting plasma may be in the form of a diffuse wind 
and may or may not have a clumpy structure.
In both cases one may hope to acquire important information from analyzing the
profiles of such lines, and to learn about the strength of the gravitational field,  distance, velocity field etc. 

Many normal, luminous stars
possess strong mass loss. 
In many cases it is possible to
derive information about the wind by analyzing the P-Cygni profiles formed in these winds. These peculiar profiles have long been known.
The first paper which contained an
explanation of P-Cygni profiles on the basis of the Doppler effect was that of ~\citet{Beals}, who gave a basic explanation of the observed line profiles from novae and Wolf-Rayet stars. Since then, these profiles have been studied both observationally and theoretically in numerous papers.
For example, \citet{Morton}, 
extensively investigated the P-Cygni ultraviolet resonant line profiles
from winds of early supergiants.

A theoretical breakthrough was made in the paper
by \citet{Sob}, in which he recognized that if there exists a velocity gradient
along the line of sight, the radiation transfer problem
becomes purely local. That is, distant regions in the flow can no longer exchange 
information using photons with frequencies
within the width of the line. The radiation which is seen at a certain frequency within a line profile by a
distant observer comes from a surface of equal line-of-sight velocity.
(see Sect. 3.1).
Following this, significant attention has been paid to computing and analyzing P-Cygni profiles and also to investigating their dependence on fundamental parameters of the flow
(see e.g. the atlas of P-Cygni profiles by \cite{CastorLamers}) .

The main goal in this paper is to calculate the line profile produced by 
a plasma which is moving in the strong gravitational field of a
compact object. 
Thus, we want to obtain an answer to the question of how much the
gravitational red-shifting can change the observed line profile in
comparison with the standard case of P-Cygni line. 
Expecting that the gravitational red-shifting may introduce some characteristic distortion to the P-Cygni profile, we wish to investigate what kind of the distortion is made and whether it can be used to deduce information about the 
wind and compact object system itself.

The plan of the
paper is as follows: in Sect.2 basic assumptions are described about
physical conditions and geometry of the line forming region. Here
the objectives and limitations of the approach are
defined; taking into account gravitational red-shifting,  in Sect.3 we first calculate a spectral line optical
depth. This is done in the Sobolev approximation which we adopt throughout the paper. 
In this section equal frequency surfaces (EFS) are
re-defined to include the gravitational red-shifting effect and the shapes of
such surfaces are calculated for several typical
velocity profiles. The results obtained
in this section are 
extensively used in the  calculation of the radiation field within a spectral
line in Sect.4 or when we numerically calculate line profiles in
Sect.5. 
At the end of the paper we provide a discussion and summarize the results.

\section{Assumptions}
As mentioned above, the problem we address in this paper may be
relevant to various astrophysical objects. 
Physical conditions in the wind driven from the accretion disk in AGN are different from 
those in the out-flowing plasma during an X-ray burst. At the same time we would like to elucidate some new features
which may arise in a spectral line profile when the influence of the gravitational field  is non-negligible.
As a zero order model we adopt the simplest
geometry and make further hypotheses about the velocity
profile, temperature distribution, opacity etc. 
We wish to consider
a minimum number of free parameters, as their overabundance will likely
obscure rather than elucidate the origin of expected new features.
We make the following assumptions
about the geometry and kinematical properties of the absorbing plasma and the source of continuum photons:

\begin{enumerate}
\item The velocity distribution is spherically symmetric.
This assumption allows a simplification to the radiation transfer problem
and also adds robustness to
the results. 
In some cases a departure from spherical
symmetry (e.g. wind from accretion disk) should be taken into
account.  For example, rotation may have some influence on the line profile. 
However, in radiationally or thermally driven accretion disk winds , the radial component of the velocity can quickly exceed 
the toroidal component, due to the conservation of the angular momentum. Notice, that this may not be the case in the MHD winds, in which poloidal velocities may not be much larger than the toroidal ones e.g. \cite{BlandfordPayne82}.
On this ground we hypothesize that inclination and geometrical effects due
to accretion disk may have more influence on the resultant line profile.

\item
The velocity profile is represented by a gradually increasing function of the radius. In the case of P-Cygni profiles from normal stars, decelerating winds were
considered in several works ( see e.g. \citet{KuanKuhi}, \cite{Marti77} ).
Some results of these papers are discussed further in the text.

\item
The only source of continuum photons is the spherical core. 
Accretion disk may influence the results in two ways: close to the BH, the disk provides strongly anisotropic radiation field; the disk may screen the emission from part of the wind. 
At the end of this paper we investigate the effect of screening
by the disk which is viewed face on.

\item
No bending of the photon trajectories is taken into account. While this effect plays an important role in formation of lines from the inner parts
of accretion discs (both Schwarzschild and Kerr) it does not play
an important role in the  "zero-order" model adopted in this paper.
The relative importance of bending can be inferred from the equation which describes the orbit of a photon in
the gravitational field of a Schwarzschild BH:

\begin{equation}\label{geod1}
\frac{1}{x}\simeq\sin{\alpha}+\frac{1}{4g_0}(3+\cos{2\alpha})\mbox{,}
\end{equation}
where $\alpha$ is the polar angle of the photon's trajectory with
the impact parameter $b$, $g_0=b/r_g$, $\displaystyle
r_g=\frac{2GM}{c^2}$ is the Schwarzschild radius, and $x=r/b$. In
the current studies we are concerned with regions of the
flow located approximately at radii $>10 r_g$. Thus,
taking $x=\infty$, $g_0\sim 20$ we obtain the
deflection angle, $\Delta\alpha\simeq 5.7^{\circ}$. Taking into
account bending of photon trajectories would introduce
a significant complication in the solution the radiation
transfer problem and would require frequency dependent Monte-Carlo
simulations of the radiation transfer in the frame of the General Relativity (GR for short), which is beyond the scope of the current studies.
\end{enumerate}

\section{optical depth}

We associate an observer situated at infinity, $O^\infty$ with the laboratory frame (''lab''- frame for short).
The fluid is moving with the velocity $v$. It is most convenient to measure absorption and emission
coefficients in the co-moving frames, $O_{\rm com}$, coinciding instantaneously with the fluid at
each point along the streamline. 
To obtain a relation between these two frames one needs to consider another local 
frame, $O_{\rm loc}$ - a Lorentzian frame which is at rest at a given point $s_0$
and
instantly coincides with the $O_{\rm com}$ frame. 
Generally $O_{\rm loc}$ and $O^\infty$ are not the same because of the presence of the gravitational field.
The
opacity in the $O_{\rm loc}$ frame is obtained according to the
transformation: $\chi^l_{\rm loc}=\chi^l_{\rm com} \,\tilde\nu/\nu_{\rm loc}$,
where $\chi^l_{\rm com}$ and $\tilde\nu$ are the absorption coefficient and the frequency of the radiation in the co-moving frame.
Notice that we neglect all effects associated with the strong gravitational field except for the change of the radiation frequency. 
In particular, we assume that  photons are propagating in straight lines,
so that $\mu_{\rm lab}=\mu_{\rm loc}$, where $\mu$ is the cosine of
an angle between the radius-vector and the direction of the ray.
The frequency of a photon that is propagating in the background gravitational field
can be found from the relation: $\nu_{\rm loc} \sqrt{g_{00}}={\rm
const}$, (\cite{LandauLifshitz}) , where $g_{00}$ is the corresponding component of the
Schwarzschild metric tensor. In the "weak field limit" it follows that
$\sqrt{g_{00}}\simeq 1+\phi/c^2$ where $\phi(r)$ is the potential
of the gravitational field.

In this paper, we adopt the pseudo-Newtonian potential of Paczynski-Wiita
(PW) (\cite{PaczWiita}): 
\begin{equation}\label{pw1}
{\displaystyle \phi(r)=\frac{GM}{r_g-r}}\mbox{.}
\end{equation}

\noindent The PW potential mimics important features of exact general
relativistic solutions for particle trajectories near Schwarzschild
black hole. This potential correctly reproduces the positions of
both the last stable circular orbit, located at $3r_g$ and the
marginally bound circular orbit at $2r_g$. This useful property, to
capture the essentials of GR effects, has been used in \citet{Dora03} to
calculate the structure of the line-driven wind near compact object, and
the results are in a good agreement with the GR
calculations of  \citet{DN} 

The Lorentz transformations between $O_{\rm loc}$ and $O_{\rm com}$ give:
${\tilde \nu}=\gamma\nu_{\rm loc}(1-\mu\beta)$, where $\beta\equiv
v/c$, $\gamma\equiv(1-\beta^2)^{-1/2}$.
Consider a photon that is emitted  at a point $s_0$.
At some other point $s$ it's co-moving frequency is
${\tilde \nu}(s)=
\gamma\nu_{\rm loc}(s_0)\sqrt{g_{00}(s_0)/g_{00}(s)}(1-\mu(s)\beta(s))$.
In order to allow for moderately high terminal velocities we will write all equations out to the second order, i.e. retain terms $\beta$, $\beta^2$, and $\phi/c^2$.
Taking into account that $\nu_{\rm loc}(s_0)\sqrt{g_{00}(s_0)}=\nu^\infty$, where $\nu^\infty$ is the frequency measured by the observer $O^\infty$, we obtain:

\begin{equation}\label{tnu1}
\tilde\nu(s)=\nu^\infty\left(1-\mu(s)\beta(s)-\frac{\phi(s)}{c^2}+
\frac{\beta(s)^2}{2}\right)\mbox{.}
\end{equation}

\noindent

The probability to emit a photon within
a frequency range ($\tilde\nu,\,
\tilde\nu+d \tilde\nu)$, and within a range of solid angles in the co-moving frame
$(\tilde\Omega$, $\tilde\Omega+d\tilde\Omega)$ is:
$dP_{e}=(4\pi)^{-1}\,\Psi(\tilde\nu-\nu_0) \,d\tilde\nu\, d\tilde\Omega$, where
$\nu_0$ is the frequency of the line in the co-moving frame, and
the line
profile function $\Psi$ is assumed to obey the normalization
condition:  $\int_0^\infty\Psi(\tilde\nu-\nu_0)\,d\tilde\nu=1$.

The optical depth between a point $s$ and $+\infty$ along the photon's
trajectory, which is a straight line in our approximation, is calculated adopting a transformation from space to frequency variable:

\begin{eqnarray}\label{tau1}
t&=&\int_{s}^\infty \chi^l_{\rm loc} \,ds'=\int_{s}^\infty\,\frac{\tilde\nu(s')}{\nu_{\rm loc}}\chi^l_{0,\rm com}\Psi \left( \tilde \nu-\nu_0 \right)\,ds'\\
&=& \int_{\tilde\nu(s)}^{\tilde\nu(\infty)}\,
\frac{\tilde\nu}{\nu_{\rm loc}}\chi^l_{0,\rm
com}\frac{\Psi\left( \tilde \nu-\nu_0 \right)}
{\left(\displaystyle\frac{d\tilde\nu}{ds}\right)_{s'}}\,d\tilde\nu\mbox{,}\nonumber
\end{eqnarray}
where $\chi^l_{0,\rm com}$ is the line-center
opacity in the co-moving frame:

\begin{equation}\label{kappa}
  \chi^l_{0,\rm com}=\frac{\pi e^2}{m c} (gf)
  \frac{N_l/g_l-N_u/g_u}{\Delta\nu}\mbox{,}
\end{equation}
where
$N_u$, $N_l$ and $g_u$, $g_l$ are respectively: populations, 
statistical weights of the corresponding levels of the line transition, $f$ is the oscillator strength of the transition, $\Delta\nu=\nu_0\, v_{th}/c$ is the Doppler line width, and $v_{th}$ is the thermal velocity.
Given the Sobolev approximation we assume that a photon, after being scattered in a line at a point $s_0$, can further interact with matter only in the immediate vicinity of this point. Due to gradients of the velocity and gravitational potential, such a photon is quickly shifted out of the resonance with the line.
We expand  $\tilde\nu$ in the vicinity of $s_0$ in Taylor series retaining two terms in $(s-s_0)$:

\begin{equation}\label{tnu2}
\tilde\nu(s) \simeq \tilde\nu(s_0)+\left(\frac{d\tilde\nu}{d s}\right)_{s_0} (s-s_0)+\frac{1}{2}\left(\frac{d ^2 \tilde\nu}{d s^2}\right)_{s_0}(s-s_0)^2 \mbox{.}
\end{equation}
Usually in the Sobolev approximation only the first order term is left
in (\ref{tnu2}), and 
in most cases this is sufficient. However, there are some pathological situations, namely a singularity in
(\ref{tau1}) in which the
second order term in the equation ~(\ref{tnu2}) is required.

Using relations ~(\ref{tau1}) and (\ref{tnu2}) and taking into account that in the Sobolev approximation the opacity is considered to be constant throughout the resonant region, the expression ~(\ref{tau1}), to the second order $v^2/c^2$, reads:

\begin{equation}\label{tau2}
t\simeq\tau_0(\mu,s)\,(1+\beta^2(1+\mu^2)-2\mu\beta-\frac{\phi}{c^2})\,\zeta(\tilde \nu)\mbox{,}
\end{equation}
where

\begin{equation}\label{tau0}
\tau_0(\mu,s)\simeq\chi^l_{0,\rm com}/ |Q(\mu,s)|\mbox{,}
\end{equation}
and

\begin{equation}\label{Qmu}
Q(\mu,s)=\alpha_1+\alpha_2 \, \Delta s \mbox{,}
\end{equation}
where

\begin{equation}\label{alpha12}
\alpha_1\equiv ({\nu^\infty})^{-1}\left(\frac{d \tilde\nu}{ds} \right)_{s}
\mbox{,}\quad\alpha_2\equiv ({\nu^\infty})^{-1}\left(\frac{d^2 \tilde\nu}{d s^2}\right)_{s}\mbox{,}
\end{equation}
and $\Delta s=s'-s$, where $s'$ is a point located on the trajectory of the photon. In practice, we take $\Delta s=const$. 
This quantity  is required in order
to evaluate numerically the optical depth in those
directions, $\mu$ at which $d\tilde\nu/ds$ vanishes (see
discussion further in this section).
The  second factor  in (\ref{tau2}), $\zeta(\tilde\nu)$ reads:
\begin{equation}
\zeta(\tilde\nu)=\int_{\nu_0-1/2}^{\tilde\nu} \,\Psi(\tilde\nu-\nu_0)\,d\tilde\nu\mbox{.}
\end{equation}
For simplicity we assume that the line absorption coefficient is zero outside the frequency interval
$(\nu_0-\Delta\nu/2,\nu_0+\Delta\nu/2)$
i.e. in terms of $\Delta\nu$, spectral line has a half thickness 1/2.
In equation ~(\ref{tau2}), it was assumed that 
$(\nu^\infty)^{-1}\simeq (\nu_0)^{-1}(1+\beta^2/ 2-\mu\beta-\phi/c^2)$.

We integrate the radiation transfer equation along characteristics, which in spherical symmetry are the rays of constant impact
parameter, $p$. 
Differentiating in a direction of $\vec{s}=z\hat{z}$, we consider
all dependent variables being either functions of  $(r,\mu)$ or ($p,z$).
Taking into account that
${\displaystyle \frac{d}{ds}=\left(\frac{\partial}{\partial z}\right)_p=\frac{1-\mu^2}{r}\frac{\partial}{\partial \mu}+\mu\frac{\partial}{\partial r}}\mbox{,}$ and $\mu=z/\sqrt{p^2+z^2}$, and $r=\sqrt{p^2+z^2}$, we find:

\begin{eqnarray}\label{dyds}
-({\nu^\infty})^{-1}\frac{d \tilde\nu}{d s}&=&\frac{\beta}{r}\left[
1-\mu^2\left(1-\frac{d\ln \beta}{d\ln r}\right)\right. \\
&+&\left.\mu \left( \frac{1}{\beta\,c^2} \frac{d\phi}{d \ln r} -\beta\frac{d\ln \beta}{d\ln r}
\right) \right] \mbox{.}\nonumber
\end{eqnarray}
Similarly, after some algebra we find:

\begin{eqnarray}\label{y0_2}\label{d2yds2}
-({\nu^\infty})^{-1}\frac{d^2 \tilde\nu}{d s^2}&=&
\left[ \frac{1-\mu^2}{r}\left( \frac{2\mu\beta}{r}+\beta'(\beta-2\mu)-\frac{\phi'}{c^2}\right)\right.\\
&+&\left.\mu\left\{ \frac{1-\mu^2}{r}\left(\frac{\beta}{r}-\beta'\right)+
\mu((\beta')^2-\mu\beta''+\beta\beta''-\frac{\phi''}{c^2})
\right\}\right]
\mbox{,}\nonumber
\end{eqnarray}
where the prime denotes differentiation with respect to $r$.

In classical Sobolev theory, ${\displaystyle d\tilde\nu/ds}$
appears to be an even function of $\mu$. 
In our case a situation is possible where for certain values of $\mu$ Doppler shifting and gravitational shifting 
cancel each other,  zeroing the right hand side of 
equation ~(\ref{dyds}).
Note that even if the $\mu \phi'$ term in ~(\ref{dyds}) is taken away the right hand side of this equation can change sign depending on whether $dv/dr$ is positive or negative if radius is increased.
Thus for { \it decelerating} flows it is possible that 
purely geometrical frequency shift, ${\displaystyle \frac{1-\mu^2}{r} {v}}$ (due to the divergence of the 
spherically-symmetrical fluid flow), can compensate for the
${\displaystyle \mu^2 {v}'}$ term.

 Strictly speaking, for those directions of $\mu$ at which
gravitational frequency shifting cancels Doppler shifting, the 
first order (in $(s-s_0)$) Sobolev approximation is formally not applicable because of the singularity found in the denominator of eq. (\ref{tau0}). 
A similar problem has been also
found by \cite{Jeffery95} who considered the relativistic,
time-dependent Sobolev approximation. In practice when numerically evaluating the integral ~(\ref{tau1}) we adopt the following
procedure: In most cases the first order Sobolev approximation works very well and only the term which is first order in
$(s-s_0)$ has been retained in ~(\ref{tau0}).
In our calculations, in most cases the singularity in  ~(\ref{tau1}) has not been detected. 
On the other hand,  in those rare situations when it was found, we have taken the second order term into account in the denominator of ~(\ref{tau2}). Doing so, we use a second order expansion (\ref{tnu2}), and
additionally specify a   
{\it constant} value of $\Delta s$. 
It should be emphasized that our
experiments in choosing different values of $\Delta s$
persuaded us that the resultant line profile is not influenced by this choice.

\begin{figure}
\includegraphics[width=180pt]{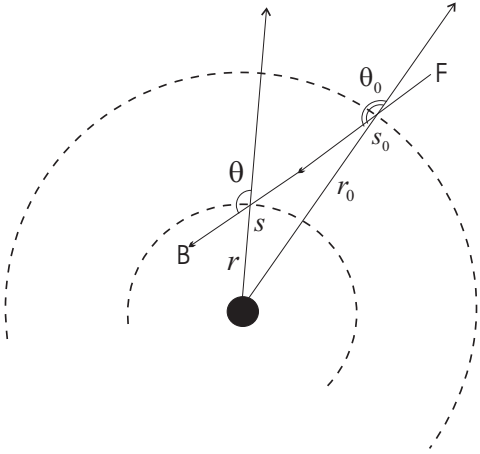}
\caption{
A photon is propagating in arbitrary direction FB from larger to smaller radii. A wind moves radially with the velocity $V(r)$. 
Corresponding projections of the radial velocity
on the direction of the ray at points $s_0$ and $s$ determine the relative Doppler shifting. 
For certain cases the latter can compensate for the gravitational shifting  between these two points (see e.g. the discussion before and after eq. ~(\ref{mu_relation})).   }
\label{illustr_2}
\end{figure}

To better understand the physical origin of the singularity,
it is instructive to consider a ray formed by photons propagating from larger to smaller radii, i.e. from $r(s_0)$ to $r(s)$ in the direction FB as depicted in Fig.~\ref{illustr_2}.
Assuming the velocity field to be spherically-symmetric and the velocity gradient positive, $dv/dr>0$, and denoting $\mu(s_0)=\cos \theta_0$, and $\mu(s)=\cos\theta$,
one can see that there are two cases to be considered: a)  $\mu(s_0)<\mu(s)<0$, so that
\begin{equation}\label{mu_relation}
|\mu(s_0)|>|\mu(s)|\mbox{,}
\end{equation} 
and b) $\mu(s_0)<0\,\mbox{,}\, \mu(s)>0$.
Rewriting equation ~(\ref{tnu1}) and keeping only first order terms we obtain:
\begin{equation}\label{dnu1}
\delta\nu/\nu_0=  - (\beta(s_0)|\mu(s_0)|\mp \beta(s)|\mu(s)|)
+\frac{|\phi(s)|-|\phi(s_0)|}{c^2}\mbox{,}
\end{equation}
where $\delta\nu\simeq\tilde{\nu}(s)-\nu_0, \tilde{\nu}(s_0)\simeq \nu_0$ and the minus sign corresponds to the case a). If $r(s)<r(s_0)$ and recalling that velocity is an {\it increasing} function of $r$, we obtain
$\beta(s_0)\geq\beta(s)$. Since
the last
term on the right-hand side of ~(\ref{dnu1}) is positive 
we see that
in both cases it can compensate for 
the term which is due to the Doppler-shift. That is,
in contrast to the case when no gravitational red-shifting is taken into account,  here a
photon emitted in the wind can again interact within the wind, thus providing a 
radiative coupling of distant points. Thus we arrive at the problem of multiple-valued equal frequency surfaces.

The problem of multiple-branched equal velocity surfaces is known
from modeling of line profiles from decelerating
atmospheres (~\cite{KuanKuhi}, ~\cite{Surdej77}).  These authors
adopted a decelerating velocity law,
$v(r)=V_{\rm const}(R_{\rm ph}/r)^l$, where $l>0$ is a parameter of
deceleration and $R_{\rm ph}$ is the radius of the photosphere. For such a velocity distribution a
coupling of distant points due to the zero relative frequency
shift is found. As a result of such coupling the radiation field at any of these points includes
not only the contribution from the source of continuum radiation ("core") but
also from distant parts of the flow.

\subsection{Surfaces of equal frequency}
\noindent
It is convenient to
introduce a non - dimensional frequency variable which measures a displacement 
of the frequency from that of the line center in terms of Doppler line width,
$\Delta\nu$:

\begin{equation}\label{ydef}
y= (\nu-\nu_0)/\Delta\nu\mbox{.}
\end{equation}
Similarly, a  co-moving version of this variable reads: 
${\displaystyle 
\tilde y= (\tilde\nu-\nu_0)/\Delta\nu 
}$, and
for the observer, $O^\infty$ we obtain
$y^\infty=(\nu^\infty-\nu_0)/\Delta\nu$ as a measure of the red/blue-shift within the line profile.
According to the Sobolev approximation 
photons of certain frequency emitted in a spherically-symmetric, 
gradually accelerated atmosphere,
may come only from the volume occupied by a thin shell of a
thickness, $l(p,z)\sim v_{th}/(dv/dr)$. 
In the limit, $\Psi(\tilde y)=\delta(\tilde y)$ this shell
becomes a surface of constant line-of-sight velocity. This is a key idea behind the calculation of the line profiles in the Sobolev approximation.
The concept of surfaces of constant line-of-sight velocity, $v
\cos{\theta}$
can be modified and expanded to account for gravitational red-shifting. 
In such a case they are be better referred 
to as equal frequency surfaces (hereafter EFS). Similarly to how the
relation ~(\ref{tnu1}) was obtained, we write:

\begin{equation}\label{yrelation}
y^\infty-\tilde y=\Delta\nu^{-1}(\nu^\infty-\tilde\nu)=\mu
u+{\Phi}-u^2/(2\zeta)\mbox{,}
\end{equation}
where $u=v/v_{\rm th}$ is the non-dimensional velocity, $\zeta= c/v_{th}$, and $\Phi=\phi/(c v_{th})$ is the non-dimensional potential.

A photon of the emitted frequency $\nu_0$
(i.e. $\tilde y=0$), emitted in the wind at a point ($z_0$, $p$), at infinity has, an observed frequency $y^\infty$. Then
an equation for the resonant surfaces reads:

\begin{equation}\label{yres}
y^\infty-\frac{z_0}{\sqrt{z_0^2+p^2}}
u\left(\sqrt{z_0^2+p^2}\right)+\Phi\left(\sqrt{z_0^2+p^2}\right)
-\frac{1}{2\zeta} u^2\left(\sqrt{z_0^2+p^2}\right)
=0\mbox{.}
\end{equation}
This equation determines the locus of the equal frequency surface (i.e. $z_0$) as a function of $p$ and $y^\infty$:
The non-dimensional form of the PW potential, (\ref{pw1}) reads: 
${\displaystyle {\Phi(x)}=\zeta/ 2(1-x
g_0)}$, where $x=r/R^*$ is the non-dimensional radius, and $R^*$ 
is the radius of the spherical core from which
the wind is launched.
A non-dimensional parameter,
$g_0= R^*/r_g$ determines the relative
importance of gravitational red-shifting (i.e. by equating
$g_0\to\infty$ one completely neglects the influence of the
gravitational field on the energy of a photon). 
Equation ~(\ref{yres}) may have multiple roots, $z_{0i}$ depending on the 
values of $y^\infty$  and $p$ and the velocity law. 
For the range of parameters relevant to this work, it is possible to show that the last term in ~(\ref{yres}) does not affect the overall properties of this equation.
It is also true that for a given 
impact parameter, $p$, and for any reasonable velocity profile $v(r)$ (i.e. smooth
single-valued {\it increasing} function of $r$), the following cases should be 
considered separately:

\begin{enumerate}
\item
Blue-shifting domain, $y^\infty\ge0$:
Equation ~(\ref{yres}) may have zero or one
root.  The latter may be located only at positive $z$ (which implies that if
$z\geqslant 0$, the superposition of Doppler blue-shifting and gravitational red-shifting can result in the observable $\nu^\infty(y^\infty)$ 
only for unique $z_0$, $p$). 

\item
Red-shifting domain, $y^\infty<0$: The
situation is more complicated: Equation ~(\ref{yres}) may have
zero, two or three roots. At $z\ge0$ there is always one
(gravitational red-shifting is stronger than Doppler blue
shifting) or no root. At $z<0$ depending on the velocity law there can be
a situation in which a superposition of gravitational
and Doppler red-shifting at some radius $r_1$ (at which gravity is stronger but the velocity
is small) can equalize 
the sum of Doppler and gravitational red-shifting at
some larger radius, $r_2>r_1$ (i.e. where gravity is negligible but velocity is high).
\end{enumerate}

\begin{figure*}
\vspace*{174pt}
\includegraphics[width=500pt]{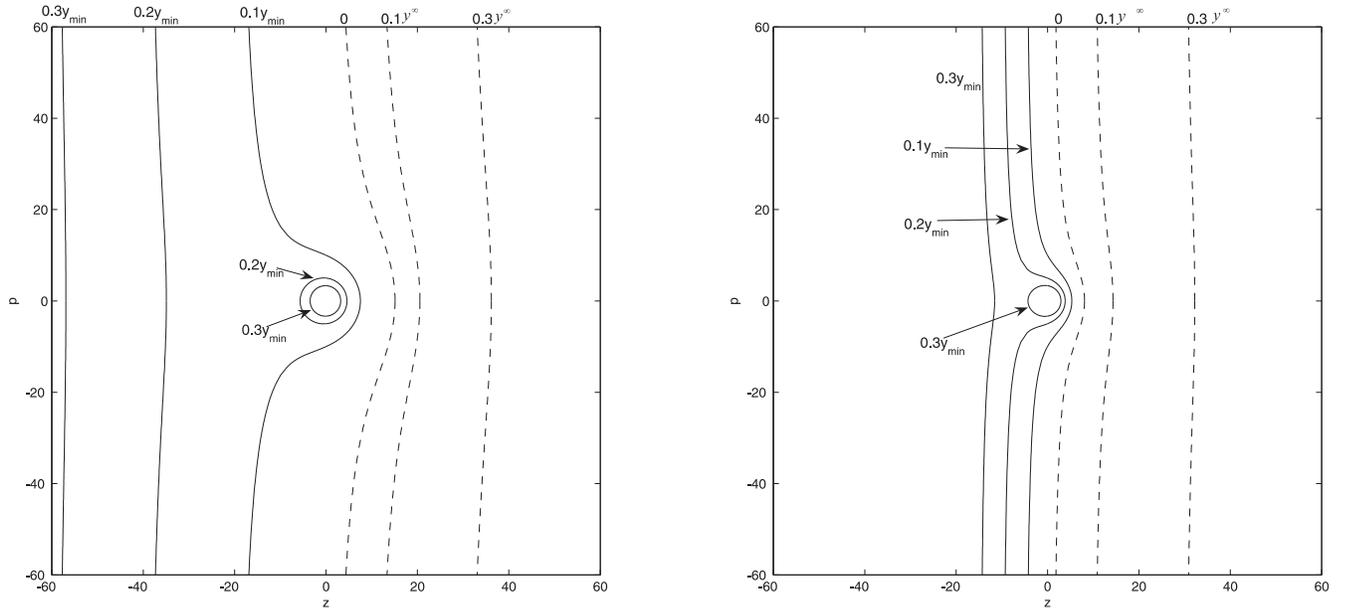}
\caption{
Equal frequency surfaces for the velocity law ~(\ref{v_lin}). 
The parameters are: $R^*=25\,r_g$ and 
$V^{\infty}=0.01\,c$
({\it left}) and  $R^*=10\,r_g$ and $V^{\infty}=0.1\,c$ ({\it right}).
Notice that a surface of equal frequency can have several branches
e.g. a ray with a given $\nu^\infty$  and $p$ can intersect it
up to three times (see text for details). The observer
is located at $z\to\infty$.}
\label{v_lin_fig}
\end{figure*}

\begin{figure}
\includegraphics[width=280pt]{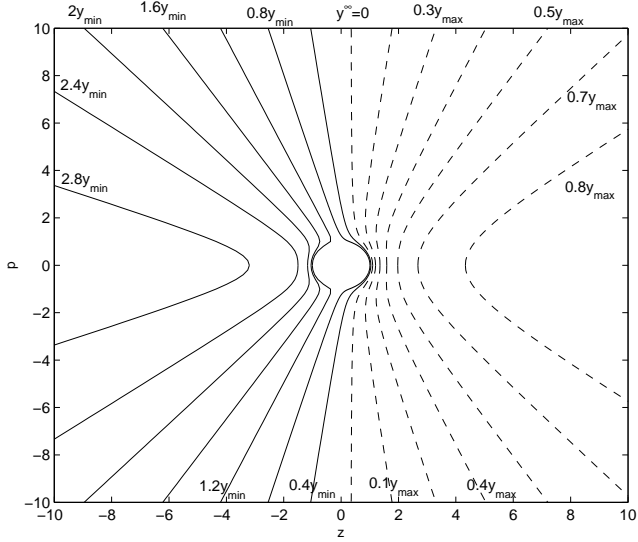}
\caption{Equal frequency surfaces for velocity law:
${\displaystyle {v}(r)=V^{\infty}\sqrt{1-\frac{R^*}{r}}}$,
where $V^{\infty}=0.1\, c$ and $R^*=15 \,r_g$. }
\label{v_sqrt_fig}
\end{figure}

\begin{figure}
\includegraphics[width=280pt]{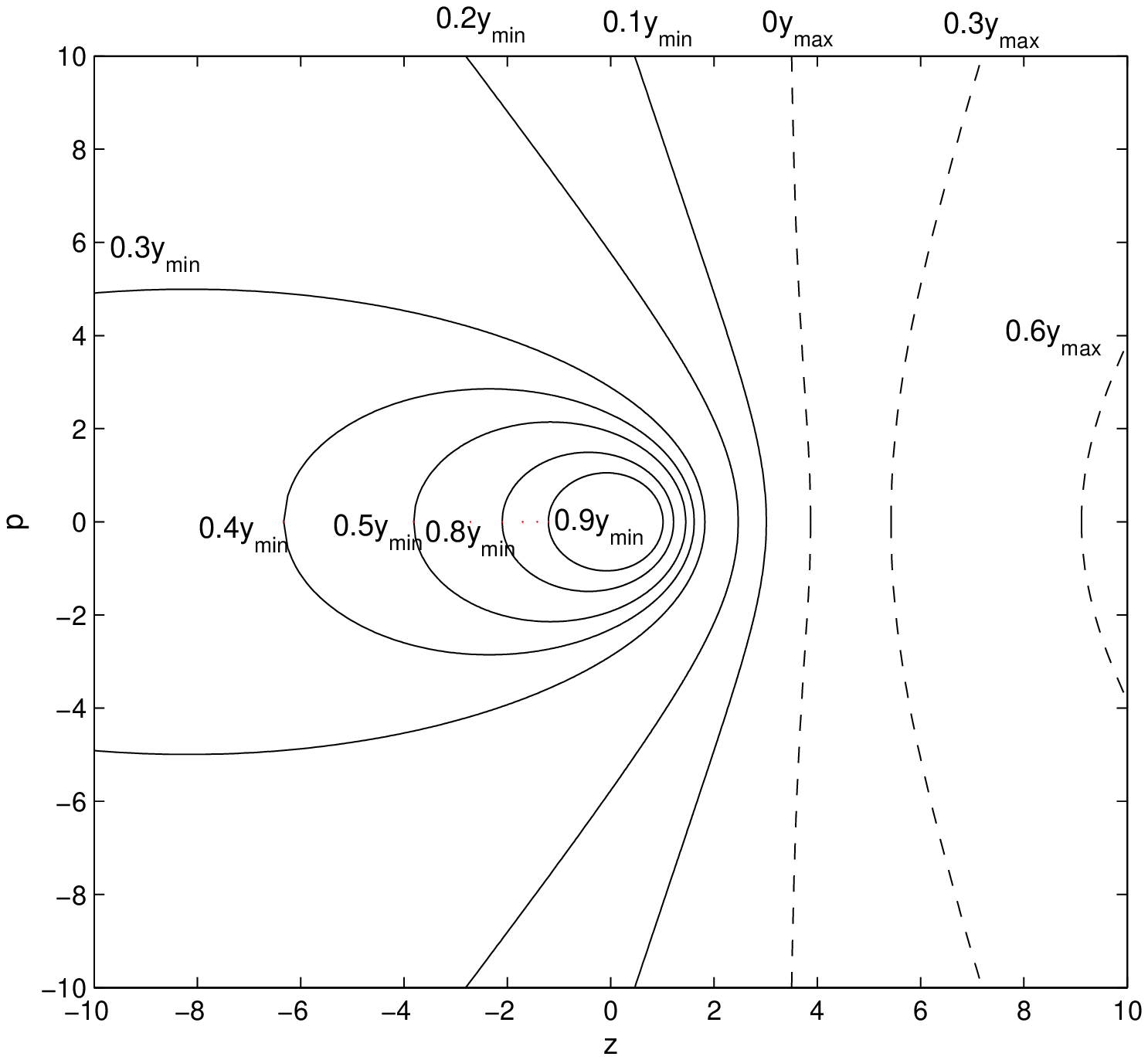}
\caption{The same problem as in Fig. \ref{v_sqrt_fig} but parameters
of the flow are different:
$V^{\infty}=0.01 \,c$ and $R^*=15 \,r_g$.}
\label{v_sqrt_fig2}
\end{figure}

The shape of resonant surfaces vary depending on the velocity profile. 
We can understand their shapes by considering only a few characteristic cases for $v(r)$. These cases are:

\begin{enumerate}
\item ${v}=0$. This case makes sense if $g_0\neq 0$, i.e. photon is 
red-shifted by the gravitational field of central object. From (\ref{yrelation}) the maximum gravitational red-shift is $y_{\rm min}=(\zeta/2)\,(1-x\,g_0)^{-1}$. For a given $y^\infty$, the EFS has the shape of a sphere of radius, $x(y^\infty)=g_0^{-1}(1-\zeta/(2y^\infty))$.

\item ${v}={\rm const}$. 
An example would be the outer part of a stellar wind in which the flow is
approaching
terminal velocity moves at almost constant speed. Alternatively, considering a spectrum from a thin, moving shell one may approximate the distribution of  velocity within such a shell as constant.
In the case of negligible gravitational red-shifting, the condition $V_z=\mu v={\rm const}$, where $V_z$ is the velocity projected on the line of sight, 
determines the locus of the EFS.
Thus, if $v={\rm const}$, the EFS has a conical shape.
An intersection of this EFS with the $p - z$ plane 
consists of two straight lines passing through the center and being symmetrical about the $z$-axis.
Even in the case of 
$g_0\to 0$ and $v={\rm const}$ there is a frequency shift along the line of sight
originating from the divergence of the flow lines in spherically symmetrical
geometry (see e.g. equation (\ref{dyds})).

\item  $v\sim{\rm const}\cdot r$. 
\noindent
Another name for this dependence is a "Hubble law",
which describes explosive events, in which
more rapidly moving particles at large radii
are outrunning the slower moving ones at small radii. 

To understand qualitatively real winds, for example those which are
quickly getting away from the potential well (for example as does a line-driven wind in O star) one may approximately consider their nonlinear velocity profile as being linear, $v\sim r$ in the inner part and $v\sim {\rm const}$ in the outer part of the flow.

\noindent
Fig.~\ref{v_lin_fig} shows equal frequency surfaces for the velocity law:

\begin{equation}\label{v_lin}
{u}(x)=U^\infty(x-1)/(x_t-1)\mbox{,}
\end{equation}
where $U^{\infty}=V^{\infty}/v_{th}$, $V^{\infty}$ is a terminal velocity,
$x_t=R_t/R^*$, and $R_t$ is the wind terminating radius, $R_t=100 R^*$. 
Thus, if
$u(x)\simeq U^\infty x/x_t$
and
$g_0\to\infty$ (i.e. no gravitational red-shifting) 
one finds from ~(\ref{yres}) that EFS is determined by the
condition $z=R_{t} \,y^{\infty}/U^\infty$. Notice that in Fig.\ref{v_lin_fig}
the equal red-shift surfaces for the
blue-shifted photons are parameterized in terms of
$y_{\rm max}=U^\infty$ (i.e the maximum blue-shift in the case of $g_0\to\infty$) . The shape of these surfaces resembles
those obtained in the case of $g_0\to\infty$.

The most important
difference from the case when $g_0\to\infty$ is the
appearance of multi-valued surfaces at the single frequency, $y^\infty$.
As was already mentioned, when $y^\infty<0$ (red-shifting),
equation ~(\ref{yres}) may have one
root at positive $z$ and two roots at $z<0$. As an example, consider a case when
$y^\infty=0.2\, y_{\rm min}$.
There are two equal red-shift surfaces: an ellipsoid-like
surface at small $r$ and a plane-like at larger radii (Fig.~\ref{v_lin_fig}, left panel). At larger
red-shift, $y^\infty=0.3 \,y_{\rm min}$ the inner surface shrinks to
smaller $r$, where the gravitational potential is stronger, while the
outer surface shifts to larger $r$, to the domain where the velocity and the corresponding
Doppler red-shift are larger.  

Terminal velocity $0.01\,c$ is 
two small to describe a continuous flow from such a small radii as $R^*=25\, r_g$. 
However, physical conditions in the flow 
(most importantly, the 
ionization state in the gas) should allow for the formation of lines from heavy ions and at the same time have enough
column density to shield these ions from being completely ionized. 
In other words, when conditions to form the line 
are just right only in a given part of a flow, and where the velocity may be approximated by the linear dependence.
For example, if the opacity peaks at small radii within a continuous flow, 
an effective maximum velocity in this shell-like region is much smaller than $V^\infty$. The low terminal velocity case is only applicable if the absorption takes place only at the low-velocity base  of the wind.

In a general case,
we expect that a transonic stellar
wind should have a terminal velocity of the order of the escape velocity at the wind launching point. For example, for the parameters in this paper, the escape velocity is of the order 0.2 to 0.3c.
Fig.~\ref{v_lin_fig} (right panel) shows such EFSs for $V^\infty=0.1\,c$.

\item 
\begin{equation}\label{vstellar}
v=V^\infty \sqrt{1-\frac{R^*}{r}}\mbox{.}
\end{equation} 

\noindent
This is a typical velocity profile for the transonic 
wind that is launched deep in a potential well.
Such a wind is characterized by a steep
transonic region, containing a critical point or multiple critical points (as in the case of line-driven winds),
and an extended plateau where the
velocity approaches ${V}^\infty$. 
This type of a velocity
law is usually considered to approximate gradually accelerated atmospheres of hot
stars. 

The characteristic behavior of the EFSs in this case can be understood 
from
the 
previously considered cases: $v\sim{\rm const}$ and $v\sim r$. 
Fig.~\ref{v_sqrt_fig} shows EFSs for the parameters $V^{\infty}=0.1\,c$ and $R^*=15 \, r_g$.
One can see that the shape of the EFS is
mostly
affected by the outer part of the flow, i.e. where $v\sim V^{\infty}$,
rather then by the inner "acceleration" part. 
The latter is more
important for the EFSs obtained for $V^{\infty}=0.01\,c$ (i.e. as in Fig.~\ref{v_sqrt_fig2}).

\end{enumerate}
Summarizing, we note that gravitational red-shifting adds new
ingredients to the classical topic of surfaces of equal frequencies:
\begin{enumerate}
\item[(a)]
Equal frequency surfaces for the red-shifted part of the spectrum ($y^\infty<0$) may be located both
in front and behind the compact object. Equal red-shift surface can break suddenly as the impact parameter approaches some critical value.
\item[(b)]
For $y^\infty<0$ the EFS may have a closed shape (ellipsoid-like) form. A ray with impact  parameter, $p$ and frequency, $y^\infty$ may
cross resonant surfaces several times depending on the velocity law and on the relative importance of gravity, i.e. proximity
to the compact object.
\end{enumerate}

Physical conditions (i.e. ionization
balance, radiation field) which affect the formation of a
particular portion of the line profile are very different on
different branches of EFS. For example, in the simple case of $v\sim r$ the absorption part is affected by the EFS which resides close to the photosphere, while the emission part is created by the EFS that may be located quite far from the photosphere (because of the large surface area of such a surface).
The same may hold true for some other velocity profiles (see section 5). 
Having said this we proceed to the next section and prepare final formulas for the
calculation of line profiles.

\section{Radiation field in a line}
The mean radiation field $J$ at any point on the Equal Frequency Surface consists of photons
emitted by atoms in the immediate vicinity of this point: a
local contribution, $J_{\rm loc}$, and a contribution from
distant points, $J_{\rm dist}$. The latter consist of photons
emitted by the core, $J_{\rm core}$, and photons which arrive from
other branches of the EFS. We
assume the intensity $I_c$ emitted by the core 
is constant over the line frequency interval
$(\nu_0-\Delta\nu/2,\nu_0+\Delta\nu/2)$, with no limb darkening taken into account. 

In the presence of 
resonant surface $\Sigma_i$ the calculation of the intensity is
straightforward. The non-local Sobolev approximation was mostly developed in the papers by ~(\cite{GrachevGrinin,RybHum1}).

\begin{figure}
\includegraphics[width=180pt]{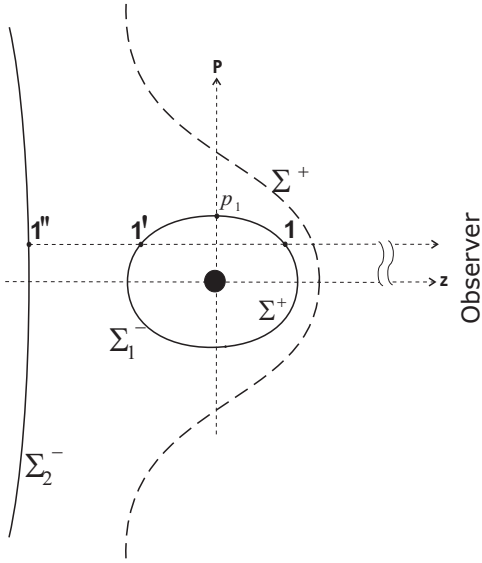}
\caption{Illustration to the geometry of equal frequency surfaces (see text for details). Not to scale. }
\label{scetch1_fig}
\end{figure}

After being emitted either by the core or
re-scattered by another EFS the intensity remains constant along
the ray $\bf s$ until it encounters the EFS at $s_i$, where it changes discontinuously according to the relation (see i.g.
~\cite{RybHum1}):

\begin{equation}\label{Intens1}
I^{s_i+0}=I^{s_i-0}e^{-\tau_{\Sigma_i}}+
S_{\Sigma_i}(1-e^{-\tau_{\Sigma_i}})\mbox{.}
\end{equation}
The situation is
illustrated in ~{Fig. \ref{scetch1_fig}}. 
In this figure $\Sigma^{\pm}$ refers to the surface of equal frequency,
and "$-$" or "$+$" sign denotes whether the corresponding part of the EFS is located at $z<0$ or at $z>0$ respectively. 
As was already established, the EFS may have  a closed or open shape. 
Thus ~{Fig. \ref{scetch1_fig}} represents a situation when gravitational red-shifting is important and there are two branches of the EFS: a closed, ellipsoid-like, and a plane-like at larger $r$ (smaller $z$), c.f. e.g. ~{Fig. \ref{v_lin_fig}}. The second possibility is that the two branches of the EFS
are both open surfaces. (To illustrate this point we draw such surface, $\Sigma^+$ with a dashed line). Other notation has the following meaning:
$\Sigma_1^-$ is a negative part of that branch which has a closed form and  $\Sigma_2^-$ is a negative and open part of the multiple-branch surface. 

The mean radiation field at the point $1$ is contributed by photons
arriving from the core and from surfaces $\Sigma^-_1$ and $\Sigma^-_2$.
Photons emitted by $\Sigma^-_1$ if not occulted by
the core,  may be attenuated by the line scattering only locally to
the point $1$: ${\displaystyle J\sim \int_{\Omega'} I_{1} (\Omega)
\beta^1_{\rm loc} \, d\Omega}$, where $I_1$ is the radiation
intensity coming from points on ${{\Sigma}^{-}_1} '$  seen from
point $1$ within the solid angle $\Omega'$ occupied by the
unobscured part of ${\Sigma^-_1} '$. One should note 
that the resonance surface which is seen from point $1$, for example, ${\Sigma^{-}_1}'$ {\it
is not}  $\Sigma^{-}_1$ but a different EFS. In a general case it should be calculated separately for
each point where $J$ is calculated. The same remains true about
all other possible branches of EFSs which can contribute to the
mean radiation field at a given point.  Generally speaking Fig.
\ref{v_lin_fig} only shows that screening is possible. The goal of taking all these geometrical
effects into account is beyond the scope of this paper.
However, when calculating the radiation
field at point $1$ one may want to take into account contributions {\it only} from
points $1'$ and $1''$, i.e. only from those points which are situated on the 
ray $(p, y^\infty)$.  Their positions remain unchanged and this fact greatly simplifies  
calculations. 
This is the first simplification in our description of the radiation field.

The second simplification arrives from the ignorance of
possible mutual screenings of the EFSs between each other and between a core (see discussion below). 
The probability for a photon to
penetrate to point $1$ without being scattered in its immediate
vicinity can be cast in the form: $\beta^1_{loc}=(1-e^{-\tau
(\Sigma^+)})/\tau(\Sigma^+)$. The intensity can be determined from the following
relation: $I_1(1')=S(1')(1-e^{-\tau(1')})$, where $S(1')$ is the
source function  at the point $1'$ on $\Sigma_1^-$. Analogously a
contribution from $1''$ can be constructed. 

Radiative coupling
between different branches of EFSs poses a difficult problem to
treat self-consistently. To overcome this difficulty different
authors adopt different approximations. 
When considering radiative
coupling between fiducial points $s$ and $s'$ ~\cite{GrachevGrinin} have
ignored the variation of the source function, assuming that
$S(s')=S(s)$. 
Another approximate approach was adopted by
~\cite{KuanKuhi} who ignored the radiative coupling between
$s$ and $s'$. In the literature this is known as the 'disconnected
approximation' (~\cite{Marti77}). 
Giving an extensive analysis of the problem of interconnection of different EFS branches for {\it decelerating} flow, these authors arrive to the conclusion that i) taking into account radiative coupling can be of importance for the calculation of the radiation force; ii) fortunately, the disconnected approximation of ~\cite{KuanKuhi} which completely neglects such coupling, works well and gives generally correct results for the source function and resultant profiles.

In this paper the
appearance of additional branches of the EFS is treated in the spirit
of the "disconnected approximation". For example, when calculating mean intensity at point
$1''$ of Fig.\ref{scetch1_fig}, we neglect any contribution from those resonant points which could be
located on the line from the core to this point. These could have important consequences because of the possible screening effect. In turn this would affect the emission part of the spectrum.
To simplify the treatment we also assume that the core serves as the only source of photons which contributes to $J$ at a given point. For simplicity we assume that the core is radiating with constant intensity $I_{c}$. 
The intensity of the radiation at $r\to\infty$ at a given impact parameter
$p$ and with a given frequency $y^\infty$ is

\begin{eqnarray}\label{Intensity1}
I^\infty(y^\infty,p)&=&I_{c}e^{-\tau(\Sigma^{+})}
+S(\Sigma^{+})(1-e^{-\tau(\Sigma^{+})})\nonumber\\
&+& S(\Sigma^{-}_1)(1-e^{-\tau(\Sigma^{-}_1)})
e^{-\tau(\Sigma^{+})}\nonumber\\
&+& S(\Sigma^{-}_2)(1-e^{-\tau(\Sigma^{-}_2)})
e^{-(\tau(\Sigma^{+})+\tau(\Sigma^{-}_1))}\mbox{,}\nonumber\\
\end{eqnarray}
where $\tau(p,y^\infty,\Sigma^{+})$ is the optical depth
on the corresponding branch of the EFS.  The notation here is the same as in Fig.\ref{scetch1_fig}: $\Sigma^{+}$ corresponds to the branch of the EFS which is located at $z>0$. Accordingly,
$\Sigma^{-}_i$ is located at $z<0$
and the subscript 1 corresponds to that branch which lies closer to the observer.

Relativistic effects may be important when calculating the escape/penetration probabilities and the source function. However, the Sobolev approximation makes
all radiation transfer purely local. 
How quickly a photon is getting out of the resonance is completely controlled and incorporated in to the optical depth. The major difference with the non-relativistic case is from the aberration effects. That is, the radiation field emitted isotropically by the core is not seen isotropic in the frame of the fluid. In the fluid frame the emission is assumed isotropic. Thus, in this frame, the expression of the source function, $S$ will have the same look (in terms of penetration/escape probabilities) as in non-relativistic case.
 In the relativistic case, the derivation of this probabilities and the of the source function is given by ~\cite{HutchSurd95}. 

Having already calculated the optical depth, (expression (\ref{tau0})) we use the arguments of 
\cite{HutchSurd95} to derive the probability of a photon emitted in a line
transition to escape in any
direction from a given point in the envelope.
To the order $v^2/c^2$ it is expressed in the form:

\begin{equation}\label{beta_loc}
\beta_{\rm esc}=\frac{1}{4\pi}\int_{4\pi} \beta^\mu_{\rm loc}
\left(1+\beta^2(3\mu^2-1)+2\mu\beta\right)
d\Omega\mbox{,}
\end{equation}
where $\beta^\mu_{\rm loc}$ denotes the escape probability in the direction $\mu$:

\begin{equation}\label{beta_mu}
\beta^\mu_{\rm loc}=\frac{(1-\exp(-\tau_0(\mu,s))}{\tau_0(s,\mu)}
\mbox{,}
\end{equation}
where $\tau_0$ is given by ~(\ref{tau0}). 
The penetration probability in the frame of Sobolev approximation reduces to

\begin{equation}\label{beta_pen}
\beta_{\rm pen}=\frac{1}{2}\int_{\mu_c}^1
\, \beta^\mu_{\rm loc}\,d\mu\mbox{.}
\end{equation}
where $\mu_c=\cos \theta_c$, so that 
${\displaystyle \theta_c=\arccos{\sqrt{1-\frac{1}{x^2}}}}$ is the maximum angle at which the core is seen from the point $x$.

Weighting the intensity by the probability of a photon to penetrate to 
a given point we obtain the mean intensity, $J$:

\begin{equation}\label{Jc1}
J=\frac{1}{4\pi} \int_{4\pi W} \,\beta^\mu_{\rm loc} (\theta,\phi) I_c \, d\Omega\mbox{,}
\end{equation}
with $\beta^\mu_{\rm loc}$ being defined by ~(\ref{beta_mu}) and $W$ is a dilution factor:
\begin{equation}
\displaystyle{
W=\frac{1}{2}\left(1-\mu_c\right)\mbox{.}
}
\end{equation}

The source function in the case of pure line scattering, in the lab frame
reads:

\begin{equation}\label{Sf1}
S=\frac{J}{\beta_{\rm esc}}\left(1+\beta^2(3\mu^2-1)+2\mu\beta\right)\mbox{,}
\end{equation}
where $\beta_{\rm esc}$ should be found from ~(\ref{beta_loc}) and 
terms of the order $v^2/c^2$ were retained in (\ref{Sf1}). In the lab frame the source function (\ref{Sf1}) is clearly anisotropic , taking higher values in the 
direction of motion.
For relativistic formula see \cite{HutchSurd95}, and for non-relativistic, e.g. \cite{Mihalas}, \cite{Castor}.
A considerable simplification is obtained in the case of a linear velocity law $v(r)\sim r$, when from ~(\ref{Sf1}) and (\ref{Jc1}), in non-relativistic case
one finds:

\begin{equation}\label{Sf2}
S_{\rm lin}=I_c \frac{\beta_{\rm pen}}{\beta_{\rm esc}},\qquad\mbox{where}\quad
\frac{\beta_{\rm pen}}{{\beta_{\rm esc}}}=W\mbox{,}
\end{equation}
Note that formulas ~(\ref{Sf2}) are applicable only in the case of a linear velocity law and when $\Phi\equiv 0$ at ~(\ref{tau0}) so that $\beta^\mu_{\rm loc}$ does not depend on $\mu$.
In the case of an arbitrary velocity law or/and if the gravitational red-shifting is taken into account the integral in ~(\ref{Jc1}) must be evaluated in its general form.

Here we again emphasize our approximation in which we ignore 
the influence of the additional branches of the EFS on the mean intensity, $J$. 
That is we ignore both negative  $\sim e^{-\tau}$ and positive
~$\sim S(1-e^{-\tau})$ contributions which may arise from resonant points on the 
line to the point where $J$ is calculated.

After the intensity, $I^\infty(y^\infty, p)$ has been calculated from (\ref{Intensity1}),
the normalized flux that is registered by the observer at infinity equals to
\begin{equation}\label{flux_inf}
F(y^\infty, p)/F_c=\int_0^{R_t}\,I^\infty(y^\infty, p)p\,dp
\mbox{,}
\end{equation}
where $F_c$ is the flux emitted by the core.

When calculating the red-shifted absorption
features, it has been found that the line
profile at some frequencies displays notable oscillations.
These oscillations are unphysical and are due to the specific behavior
of the source function on $\Sigma^+$ (see
~Fig.~\ref{scetch1_fig}). The same effect has been found by
~\cite{Marti77} for a decelerating wind from a normal star. 
We find that the amplitude of these oscillations is reduced (very slowly in our
calculations) when taking more points ${p_i}$. This is in
fact a numerical artifact which is related to the fact that the brightness
of $\Sigma^+$ strongly depends on $p$ when $p$ approaches its
maximum value $p_1={\rm max}(\,p\,\,\mbox{on }\,\Sigma^+)$. Thus,
prior to calculation of the integral ~(\ref{flux_inf}) we
calculate $p_1$, -the intersection of $\Sigma^+$ with $z=0$ plane.
Then we split ~(\ref{flux_inf}) into $\displaystyle
\int_0^{p_1}I^\infty p\,dp + \int_{p_1}^{p_{out}}I^\infty p\,dp $.
This procedure allows us to eliminate completely spurious jumps of $S$ on the
boundary of $\Sigma^+$. It can be also useful in the calculations of decelerating winds from normal stars where 
there are EFSs with sharp boundaries.

\section{Calculation of line profiles}
Before we proceed to closer examination of line profiles we need to make several assumptions about physical conditions existing in the flow. 
It is assumed that a spherically-symmetric wind originates
at the photosphere which emits radiation in continuum. The radius of the photosphere is $R^*$.
The radiation emitted by the core is resonantly absorbed (scattered) in a line of
rest frequency $\nu_0$ in the moving plasma. The relative importance of gravitational and Doppler red-, blue-shifting is controlled by parameters  $R^*$ (i.e. $g_0=R^*/r_g$) and  $V^{\infty}$, respectively. Additionally, we specify the distribution of the opacity. Following the recipe of ~\cite{CastorLamers}, we
parameterize the radial optical depth $\tau_r=t(\mu=1)$ (c.f. eq. (\ref{tau2})), as a function of the velocity. The following dependencies are considered:

\begin{equation}\label{tau_law1}
\tau_r(w)=T_0(k+1)\frac{(1-w)^k}{(1-w_c)}\mbox{,}
\end{equation}
and 

\begin{equation}\label{tau_law2}
\tau_r(w)=T_0(k+1)\frac{w^k}{(1-w_c^{k+1})}\mbox{,}
\end{equation}
where $w\equiv v/V^{\infty}$ is the non-dimensional velocity,  $w_c=v(R^*)/V^{\infty}=0.01$ and $k$ is a free parameter.
The parameter $T_0$ is related to the total optical depth at the line center.

\begin{figure}
 \includegraphics[width=500pt]{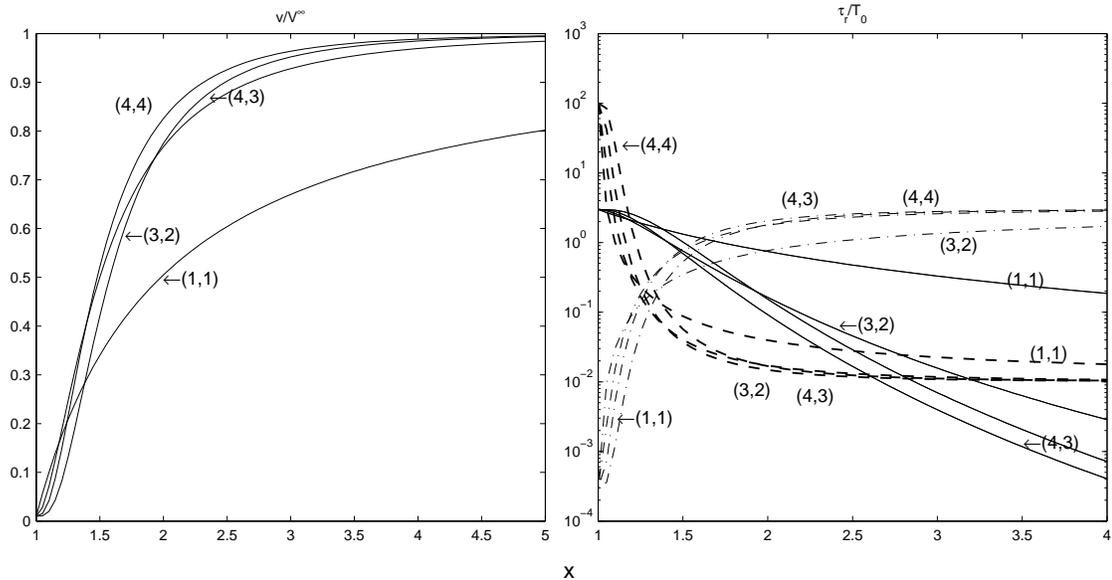}
 \caption{Velocity (left) and opacity (right) laws. Velocity law (\ref{vstell_gen}) with $w_c=0.01$. 
Opacity curves: solid line: equation (\ref{tau_law1}); dashed: equation (\ref{tau_law2}), k=-2;
dot-dashed: equation (\ref{tau_law2}), k=2.
Curves are marked by pairs of $\alpha_1$,$\alpha_2$ from
(\ref{vstell_gen}).
}
\label{veloc_opac}
 \end{figure}

We adopt the following velocity laws: the linear velocity law (\ref{v_lin}), and the ''stellar'' type velocity law:
\begin{equation}\label{vstell_gen}
w=w_c+(1-w_c)\left(1-\frac{1}{x^{\,\alpha_1}} \right)^{\alpha_2}\mbox{,}
\end{equation}
where $\alpha_1$, $\alpha_2$ determine the slope and shape of the velocity profile.
The velocity and opacity distributions for different pairs of $\alpha_1$ and $\alpha_2$, and for $k=2$ and $k=-2$, are shown in Fig.~\ref{veloc_opac}.

For a given $g_0=R^*/r_g$, we specify $y^\infty$ and calculate $v(r)$
from (\ref{v_lin}) or (\ref{vstell_gen}), and then calculate equal frequency surfaces from (\ref{yres}), then calculate
$\tau_r$ from (\ref{tau_law1}), (\ref{tau_law2}), and the source function from
(\ref{Sf1}), and then the spectrum from (\ref{flux_inf}).

\begin{figure*}
\vspace*{174pt}
\includegraphics[width=500pt]{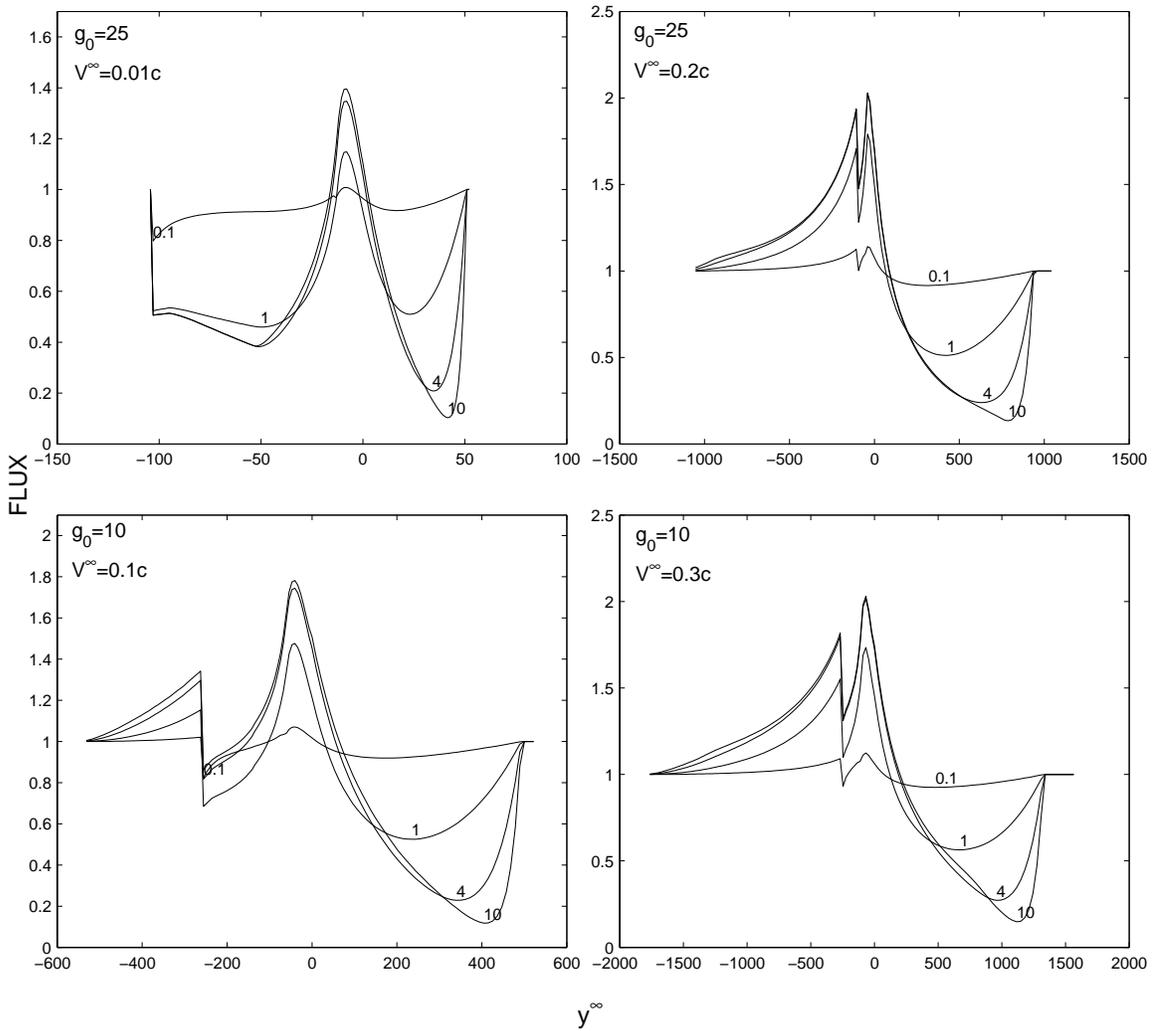}
\caption{
Profiles for $R^*=25 r_g$, $V^{\rm \infty}=0.01\,c$ (upper left),
$R^*=25 r_g$, $V^{\rm \infty}=0.2\,c$ (upper right),
$R^*=10 r_g$, $V^{\rm \infty}=0.1\,c$ (lower left),
$R^*=10 r_g$, $V^{\rm \infty}=0.3\,c$ (lower right),
and the linear velocity law (\ref{v_lin}). Curves are labelled by the total optical depth $T_0$. Notice the different vertical and horizontal scales in different panels. The left hand plots have $V^\infty<V_{esc}$, where $V_{esc}$ is the escape velocity at the wind launching point.
Vertical axes: flux. Horizontal axes: frequency shift, $y^\infty$.
}
\label{prof_lin1}
\end{figure*}

\begin{figure*}
\vspace*{174pt}
\includegraphics[width=500pt]{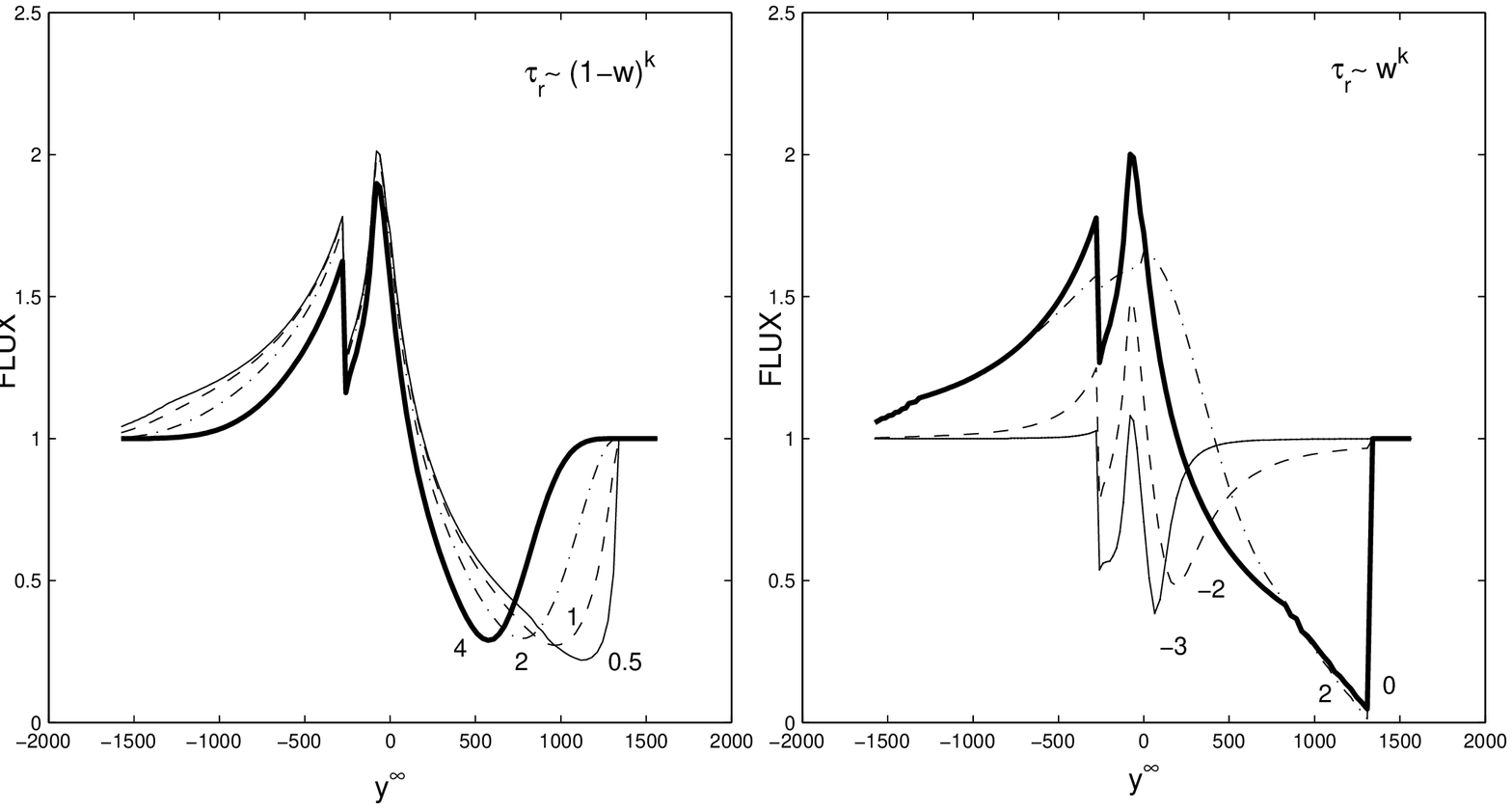}
\caption{
The effect of the distribution of different absorption laws: $\tau_r\sim (1-w)^k$ (left) and 
$\tau_r\sim w^k$ (right).
Profiles for $R^*=10 r_g$, $V^{\rm \infty}=0.3\,c$ and $T_0=4$ 
and the linear velocity law (\ref{v_lin}).
Curves are labeled by the parameter $k$. 
Vertical axes: flux. Horizontal axes: frequency shift, $y^\infty$.
}
\label{prof_lin1a}
\end{figure*}

Given the complex shape of the EFSs, i.e. depending which branch of the EFS is tracked when looking for the roots of the equation ~(\ref{yres}), we switch between $p$, $z$ or $r$, $\mu$ as 
independent variables.
Additionally, in all calculations presented in this paper the occultation by the core is taken into account. Other parameters of the model are 
$v_{th}=5.7\cdot 10^6\,{\rm cm\,s^{-1}}$, $R_t=100\,R^*$.
The results are shown in ~Fig.~\ref{prof_lin1} - ~Fig.~\ref{prof_stel3} and we consider them in turn.

\subsection{$v\sim r$ law}
In Sect. 3 it has been established that in the case of  linear velocity law (\ref{v_lin}), the EFS (the red-shifted part) may have two disconnected branches if
the influence of strong gravity is taken into account and provided parameter $g_0$ is small enough (c.f. Fig. \ref{v_lin_fig}).
This is easily understood from the following arguments:
in a normal stellar wind
$g_0=R^*/r_g\to\infty$, and  the absorption feature may be formed only in that part
of the wind which is approaching the observer, and thus is observed as blue-shifted.
In case of a static configuration 
($v=0$) the EFS has the shape of a sphere of radius $x=\vert \zeta/2g_0 y^\infty \vert$. A segment of this sphere which is
projected on the core forms the {\it red-shifted} absorption line.
Red-shifted photons could come both from the receding  part of the wind which is behind the core, and also from that branch of the EFS which is in front of it. In some cases, the re-emission from the receding gas may be faint since the source function is small: $S\sim r^{-2}$,
and we may expect to observe a red-shifted absorption
feature. In such a case the resultant profile will be characterized both
with blue- and red-shifted absorption features. 

Fig. (\ref{prof_lin1}) shows profiles for different terminal velocities and different launching radii.
The strongest second absorption is observed for the lowest terminal velocity
$V^{\infty}=0.01\,c$ (upper left).
Consider it in more detail:
the maximum width of the red-shifted absorption is set by $-y_{\rm max}$ and $y_{\rm min}$; in the picture, these are: $y_{\rm max}=52.2$ and $y_{\rm min}=-104.4$.  We can see that the width of the second absorption component is of  gravitational origin. Parameter $V^{\infty}$ determines the width and strength of the blue-shifted emission component. 

Profiles for larger terminal velocity  $V^{\infty}=(0.1-0.3)\,c$ and different $g_0$ show that for larger $V^\infty$ and for a given $y^\infty<0$, the EFS is located closer to the core and scatters more radiation, smearing the red-shifted absorption line.
A peculiar characteristic of these profiles is a red-shifted absorption line superimposed on the background of the red-shifted emission component. 
For example, if  $g_0=10$, $V^\infty=0.1c$, (lower left),
the red-ward edge of the emission is set by $-y_{\rm max}=-522$ and the  red-ward edge of the red-shifted absorption by $y_{\rm min}=-261$. 

In Fig.\ref{prof_lin1a} is shown the effect of different distributions of the opacity.
For $\tau_r\sim (1-w)^k$ (left), resultant profiles
are generally similar to those of $k=2$ and $\tau_r\sim w^k$ (right). However in other cases the results are quite distinct for different $k$.

\subsection{More realistic ${v(r)}$}

\begin{figure}
\includegraphics[width=500pt]{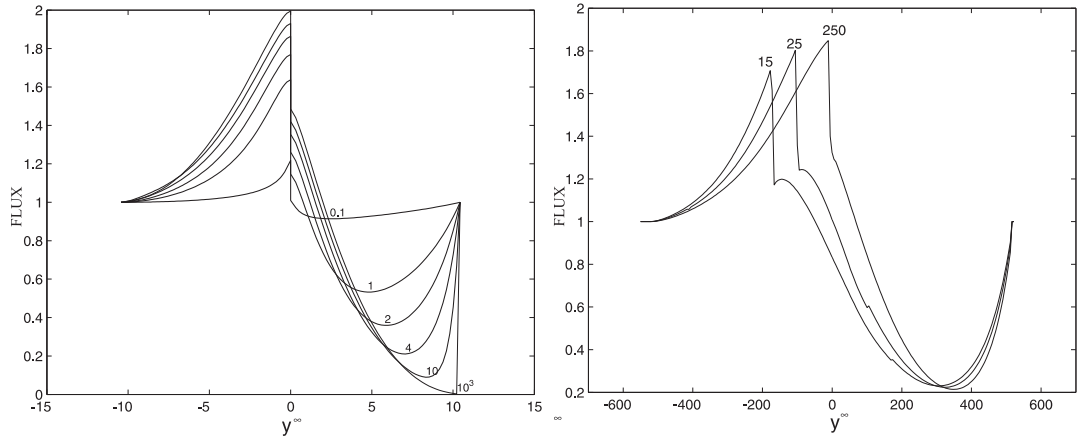}
\caption{
Profiles for $v(r)=V^{\rm \infty} \left( 1-\frac{R^*}{r} \right) $.
On the left the wind is launched at
$R^*=25\cdot 10^3\, r_g$ , having $V^{\rm \infty}=1897\,{\rm km\, s^{-1}}$.
Each curve on this panel is labeled by the total optical depth, $T_0$.
On the right the wind has $V^{\rm \infty}=0.1\,c$ and $T_0=4$, and curves are labeled by the parameter $g_0=R^*/r_g$.
In all cases the
optical depth $t\sim
(1-w)$. }
\label{prof_stel1}
\end{figure}

First we 
calculate line profiles for parameters of the flow relevant to those of a wind from a normal star. These provide a good test of our results against those of ~\cite{CastorLamers}. 
Results are shown in Fig.\ref{prof_stel1}.
They are calculated for different values of $T_0$. The terminal velocity is
$V^{\infty}\simeq 1897\,{\rm km\, s^{-1}}$.

The parameter $g_0=25\cdot 10^3$, is taken large enough to provide that gravitational red-shifting is completely negligible in this case.
Thus one expects to see no deviation of the resultant profiles from those of P-Cygni. The
profiles shown in
~Fig.~\ref{prof_stel1} are in good agreement with those presented
in ~Fig. 7 of ~\cite{CastorLamers} (note, however that these authors use $v/V^{\infty}$ as a frequency variable).
In this case, the wind approaches the terminal velocity much more quickly than in the previous two cases.
Profiles for the velocity law ~(\ref{vstellar}) are shown in Fig. (\ref{prof_stel1}) (right) for $T_0=4$. 
The increasing influence
of the gravitational red-shifting (decreasing of $g_0$) results in shifting of the edge between the emission and absorption components
to the left (to larger red-shifts). 
For example, at $g_0=15$, the position of the edge $y^{\infty}\simeq -155$ is roughly determined 
by the $\Sigma^+$ component of the EFS (c.f.  Fig. \ref{v_sqrt_fig} at $y^{\infty}=0.8\,, y_{\rm min}=140$). 
An interpretation is that the red-shifted absorption almost exactly compensates for the red-shifted emission.
Increasing $g_0$ from 15 to 250, results in a shifting of this edge to the right, finally producing 
line profile of the type shown in ~Fig.~\ref{prof_stel1} (left).

As in the case of linear velocity law,
different distributions of opacity can significantly change the results (e.g. \citet{Castor} ). The opacity is modified by varying the parameter $k$ in ~(\ref{tau_law1}),(\ref{tau_law2}). Additionally, parameters $\alpha_1$ and $\alpha_2$ can also be varied if using a general form of the velocity law (\ref{vstell_gen}).

\begin{figure}
\includegraphics[width=500pt]{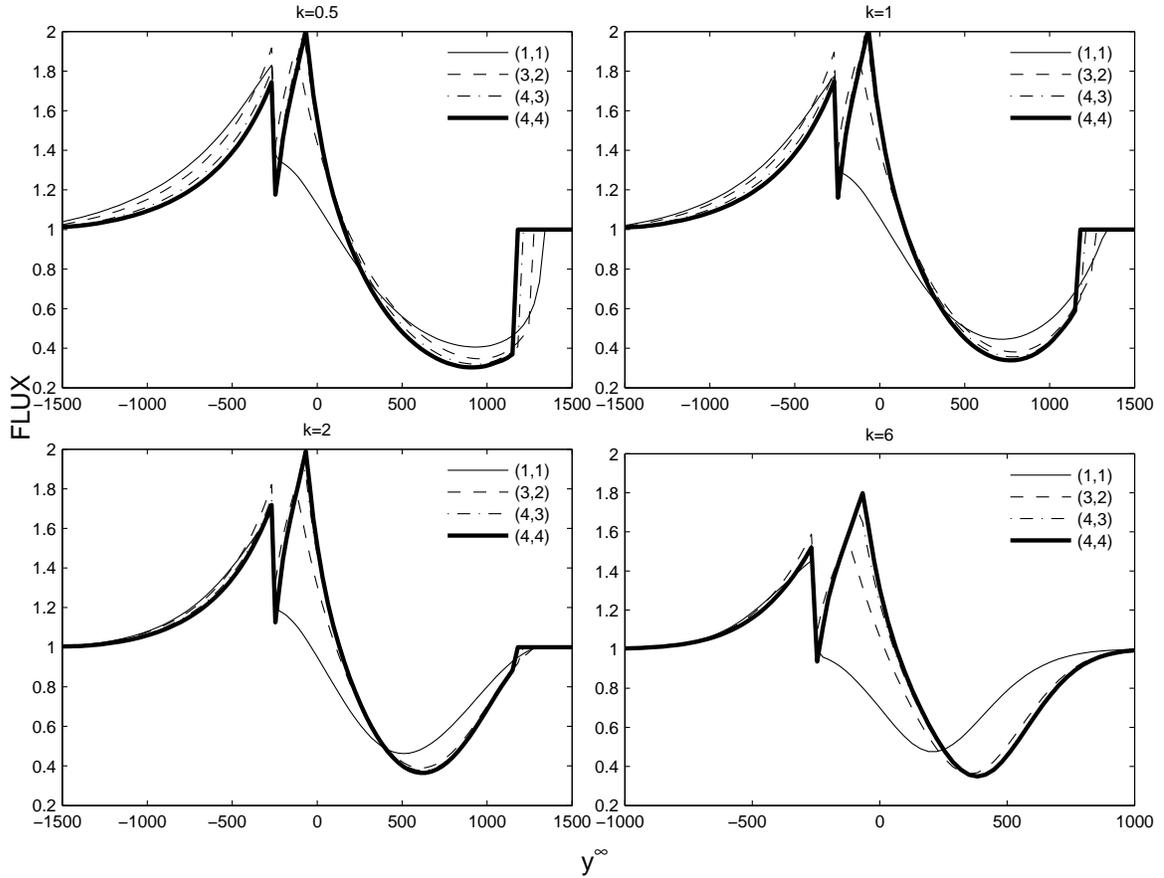}
\caption{
Profiles for $R^*=10 r_g$, $V^{\infty}=0.3\,c$, $T_0=3.$
The velocity law (\ref{vstell_gen}) and 
$\tau\sim(1-w)^k$. 
Legend; curves are labelled by pairs of parameters $\alpha_1$, $\alpha_2$.}
\label{prof_stel2}
\end{figure}

A set of profiles for $\tau\sim(1-w)^k$ is shown in Fig.~\ref{prof_stel2}.
The profiles are calculated for $R^*=10 r_g$, $V^{\infty}=0.3\,c$, and 
$T_0=3$.
Each panel shows them for different pairs of $\alpha_1$, $\alpha_2$ for the particular value of $k$. Again for $\alpha_{1,2}\neq 1$ a narrow absorption line is superimposed on the broad blue-shifted emission line. Notice, that in Fig.~\ref{prof_stel2}, the slowly accelerating wind ($\alpha_{1,2}= 1$) shows more absorption, than in other cases of steeper accelerated winds. The center of gravity of the emission peak in this case is dominated by photons coming from EFSs which are located quite far away, behind the core (wind is accelerating to slowly). These surfaces reflect too few photons ($S\sim 1/r^2$) and as a result, the gravitationally red-shifted absorption line dominates over the red-shifted emission.

\begin{figure}
\includegraphics[width=500pt]{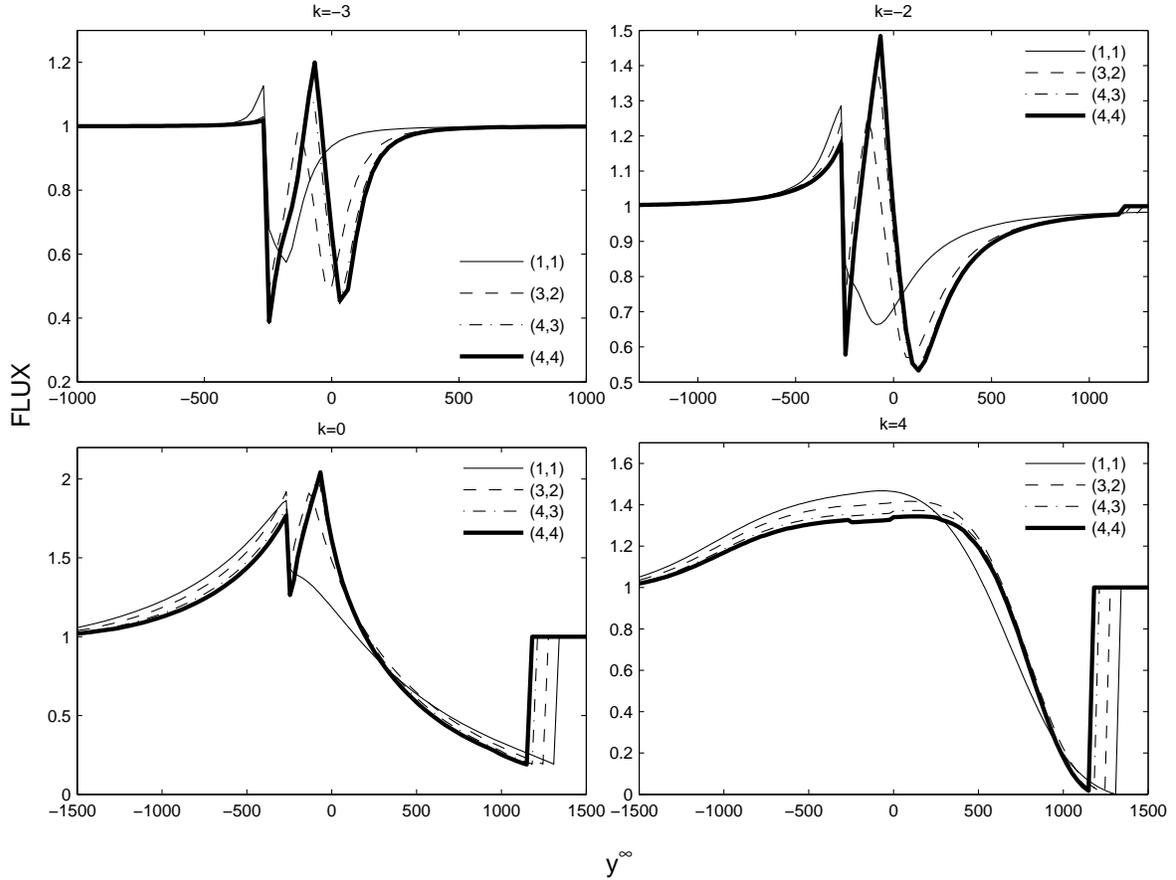}
\caption{
The parameter setup and labeling of the curves are the same as in Fig. \ref{prof_stel1}. The opacity distribution:
$\tau\sim w^k$. }
\label{prof_stel3}
\end{figure}

Fig. ~\ref{prof_stel3} shows the results for the same set of parameters as before but now for $\tau\sim w^k$. We see that the results are considerably different from the previous case. Only for $\alpha_{1,2} =1$  are there profiles which look like classical P-Cygni profiles. In most cases, the narrow absorption is superimposed on the blue-shifted emission. For $k=-3$ and $k=-2$, profiles look similar to those obtained from the linear law (c.f. Fig.~\ref{prof_lin1}).

\subsubsection{Shielding by the disk}
\begin{figure}
\includegraphics[width=500pt]{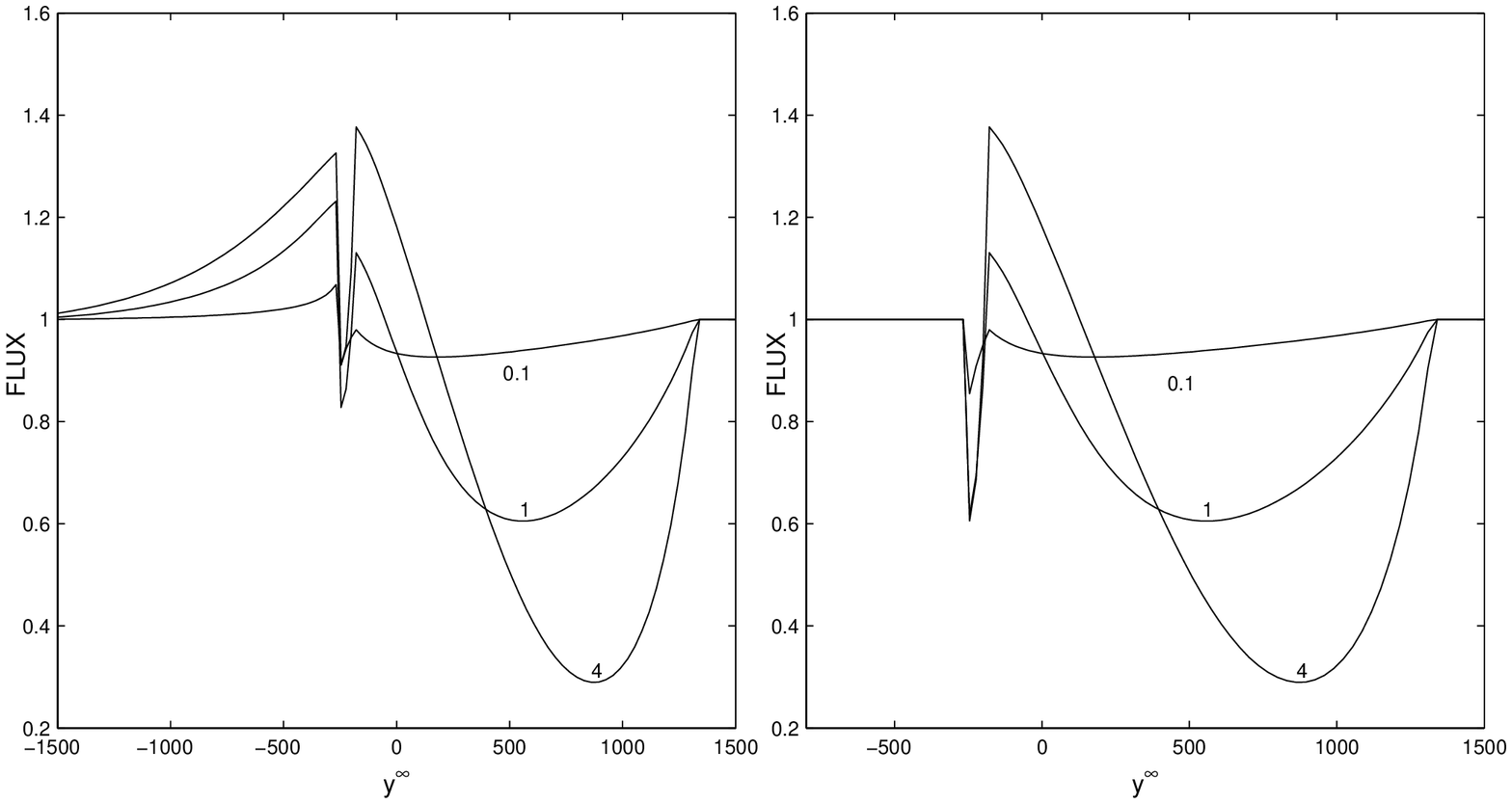}
\caption{
The effect of shielding of the emission component. 
Curves are calculated for $\tau\sim (1-w)$; left: $\tau_{\rm sh}=1$; right: $\tau_{\rm sh}=10$; labeling: total optical depth $T_0$. Other parameters: $R^*=10r_g$, $V^\infty=0.3c$.}
\label{prof_stel4}
\end{figure}

Our method, as presented in this paper, does not allow to consider 
anisotropic radiation field
of an
accretion disk or some additional attenuation of the emission. However, we can gain some insight by considering a wind that is viewed face on (i.e. perpendicular to the disk plane) and accounting for the blocking of those photons which are coming from behind such a disk. 
Since a significant part of the emission profile is formed by these photons, we
want to know what happens if this emission is reduced.
Thus we attenuate those photons by a factor of $e^{-\tau_{\rm sh}}$, where $\tau_{\rm sh}$ is the attenuation optical depth.
The results are shown in Fig. \ref{prof_stel4}. As may be expected from the previous results,  suppression of the emission leads to absorption-emission-absorption profiles regardless of the velocity law. That is because the emitting branches of the EFS, i.e. $\Sigma_1^-$,
$\Sigma_2^-$  in Fig. \ref{scetch1_fig} are screened but the absorptive branch
$\Sigma^+$ is still there and absorbing radiation.

\section{Discussion}

The strongest double absorption profile is found for the $v\sim r$ velocity
law.
The superposition of
the gravitational red-shift and Doppler shift easily produces two pronounced absorption features if the velocity
law is not steep (as in the case of linear law) or the maximum velocity within the line forming region is small enough (as in ~Fig.~\ref{prof_lin1}, upper left). 
More realistic velocity law (\ref{vstell_gen}) also gives W-shape profiles for
several different values of $\alpha_1$, $\alpha_2$.
In case of a greater terminal velocity $v=0.1-0.3\,c$ (i.e. of the order of the escape velocity at the base of the wind) in most cases the 
absorption feature is superimposed on the red emission wing of the line
for both $\tau_r\sim (1-w)^k$ and $\tau_r\sim w^k$ distributions of the opacity. For the latter distribution, the W-shape profile is most pronounced, sometimes having approximately equal absorption troughs in red and blue part of the spectrum (Fig.\ref{prof_stel3}). Blocking of the emission (as in the case of an obscuring disk) also results in W-shape profile (see Fig.\ref{prof_stel4}).

In order to produce an absorption feature, a certain resonant
surface (or surfaces) must reside on the line of sight between the
source of photons and an observer at infinity. In the case of the stellar
(low gravity) wind and provided the velocity is gradually increasing, each EFS is
single-valued. On the other hand, if the wind is decelerating, the EFS is not
unique. If this is the case then photons emitted by one branch of the
EFS can be scattered by the other and vice-versa. As was
shown above, the gravitational field introduces several new
aspects, changing both the shape and the locus of EFSs.
The
most important difference is the appearance of the resonant surface in front of the core and which is located in the
{\it red-ward} part of the frequency domain.
This component of the absorbing surface intersects the
line of sight to the core and scatters any line radiation that crosses it. While multi-branched EFSs are not
new, the appearance of the surface of equal red-shift in front
of the core for outwardly accelerated flow 
is not known in the theory of stellar winds. 

This may be made clear
if we imagine a static and extended
distribution of plasma around a compact object.
In this imaginary exercise, there is only gravitational red-shifting which is important
and the corresponding EFS is just a sphere that is concentric with the
center of gravity. 

In a moving plasma this spherical EFS is
transformed depending on the velocity law such that for 
certain red-shifts there is a component in front of the the core (c.f. Fig. (\ref{v_lin_fig})).
Absorption within this EFS
may form an absorption feature in
the red-shifted part of the line profile.

The emission component is formed in the same way
as in stellar case, namely because of the large surface area of
those EFSs which are not intersecting the line of site between the
observer and the core. 
This red-shifted emission is superimposed on the absorption line.
The net result depends on the distribution of the opacity, i.e. on the amount of scattered radiation.
The blue-shifted part of the profile is
formed because of the interaction of the {\it blue} continuum
photons with resonant EFSs giving the absorption feature for
the blue part of the profile. 
The value of $V^\infty$ roughly sets the width
of the red-shifted emission and blue-shifted absorption.
Results show that the centroid energy of the blue-shifted absorption depends,  on the distribution of the opacity, and 
can be seen at significantly smaller shifts.
If the gravitational red-shift effect is significant but smaller than the pure Doppler effect, there exists some maximum red-shift at which the 
EFS has a component at $z>0$. This component attenuates the core radiation as $I_{c}e^{-\tau(\Sigma^{+})}$ ~(\ref{Intensity1}), 
and produces an absorption feature superimposed on the emission component, as in Fig. \ref{prof_stel2}-\ref{prof_stel4}.. Here, again, the details depend sensitively on the distribution of the opacity.

The intensity $I(y^\infty,p)$ is composed of
the radiation of the core, $\sim I_ce^{-\tau}$ and the contribution
${\displaystyle \oint_{\Sigma_i} \,
S(x,y)(1-e^{-\tau_{\Sigma_i}})\, dx \,dy}$, where the integral is
carried along the resonant surface. In the case of pure line scattering:
$S\sim 1/r^2$, and the "brightness" of the emission line is determined by the 
surface area of the resonant surface, balanced by the $1/r^2$ divergence of
the radiation flux.  
Thus, it is not enough to have opacity concentrated to smaller, or larger $r$.
If at small $r$ the absorption and gravitational field are strong the absorption feature may be strong but the red-shifted emission formed by scattered radiation is also strong. 
In the other extreme, at large $r$ the emission may very faint because of the 
small flux.
The shape of the EFS is also important. Without actual calculation of the EFS there is no way to tell what is stronger: the red-shifted emission or red-shifted absorption.

The
conclusion is: if absorption takes place in a spectral
line within a wind provided that the plasma is moving radially and gradually accelerated in strong
gravity, a profile with two absorption features is quite possible; in some cases there is a prominent emission
component, and the profile is observed as W-shaped.

The results are
sensitive to the assumptions about the opacity and velocity;
the geometry of the
EFSs depends on the distribution of $v(r)$ and on $g_0=R^*/r_g$; the amount of scattered radiation depends on 
the distribution of the opacity. If it is concentrated towards the star (as in Fig. \ref{veloc_opac}), the conditions favor the formation of a W-shape profile. However in each case different possibilities in choosing $v(r)$ and $\tau_r(r)$ should be considered.

\section{Conclusions}

Our goal in this paper has been to study shapes of spectral lines from plasma which is rapidly
moving in the vicinity of a neutron star or a black hole. The
latter case can account for both stellar (Galactic Black Hole Candidates) and supper-massive BHs (AGN). In the well studied
case of winds from normal hot stars, a rapidly moving wind interacts
with the continuum radiation of a star and produces
a P-Cygni profile. From the theory of stellar winds it is known that in the case of an arbitrary spherically-symmetrical
distribution of plasma which is moving with a gradually increasing velocity, an absorption line
that is blue-shifted with respect to the emission line is
observed. This is widely interpreted as a fingerprint of moving
plasma in a variety of astrophysical situations. 

In this paper,
shapes of line profiles for several velocity and opacity distributions 
were calculated taking into account gravitational red-shifting.

Strong gravitational red-shifting helps the radiation to escape efficiently and to interact with matter only locally. This was already established to be dynamically important.
The papers by \citet{Dora03} and \citet{DN} addressed a problem of plasma acceleration
driven by radiation pressure in spectral lines provided that the flow
is launched in the vicinity of a compact object. It was shown that it is important to include gravitational red-shifting in the calculations of the radiation force. The radiation pressure depends on $d\phi/dr$ and $dv/dr$, while in the case of O-type star wind it depends only on $dv/dr$. Proximity to compact object, i.e. strong ionizing radiation, makes it difficult for heavy ions
to survive against being completely stripped of electrons, so the efficiency 
of this additional mechanism in real situation depends on the simultaneous solution for the ionization balance and  accounting on other possible mechanisms (such as clumping) to prevent over-ionization of the flow.

We are also motivated by the current accumulating 
evidence for the gravitationally red-shifted narrow absorption features
in many AGN spectra 
as well as by
observations of gravitationally red-shifted absorption
lines in the X-ray burst spectra of neutron stars.
Potentially interesting example of the latter is presented by the observations of the gravitationally red-shifted absorption lines in the X-ray burst spectra of EXO0748-676 (\citet{Cottam}).
If the interpretation is correct, these features are produced close to the compact object, where extreme conditions are coupled with high amplitude fluctuations of the radiation field and shortest dynamical time scale. 
As a result, these lines may be highly variable or/and transient.
The non-detection of gravitationally red-shifted lines in the observations of the other good looking candidate GS, 1826-24 (\cite{Cottam2}) may be an example.

Given the diverse nature of these objects,
it is desirable to consider a model with a minimum number of free parameters
and in so doing we assume our outflow to be 
spherically-symmetric and exposed to the continuum radiation
of a spherical core. 

Realistic models should consider a departure of the outflow and 
the radiation field from 
spherical symmetry. Bending of the photon's trajectories in the strong gravitational field of the compact object, may play some role. In our calculations we did not take into account because we were concerned 
with regions of the flow located approximately at radii $> 10 r_g$ .

In our calculations the idea of the equal frequencies surfaces 
(EFS), plays a major role. 
The strong gravitational field changes the shape and locus of such a surface. Their topology is complex; the gravitational field strongly distorts the P-Cygni profile. 
Some branches of the EFSs in the {\it
red-shifted} part of the spectra are found to be in front of the
core, meaning the possibility for the absorption component to be
observed as {\it red-shifted}, with respect to the emission.

From numerical calculations, which are second order accurate in $v/c$, 
it is established that a superposition of Doppler and
gravitational shifting of frequency can distort the P-Cygni profile in such a way that 
blue- and red-shifted absorption features are observed simultaneously.
Often the red-shifted absorption line is superimposed on the emission wing.
This effect is, of course, strongest close to the 
compact object. However, the emission part is also stronger there, and it may smear the absorption trough. 
Necessarily, the
second absorption
arises if the velocity profile is not very steep and
at the same time the line forming region is situated within
several tens of $r_g$. However the former requirement is not crucial,
if the the line opacity peaks close to the compact object. 
Profiles with more than one emission and two absorption features are possible.
However, in a model of pure line scattering, those additional (i.e supplementary to W-shape profile) features would be
considerably weaker.

Different modifications of  a ''stellar'' type velocity laws were adopted. Such velocity profiles describe stationary spherically-symmetric hydrodynamical flows.  The well known example of such is the $v\sim\sqrt{1-1/x}$ profile. Note
that different modifications of this law were obtained by different authors using  both theoretical and observational arguments (see e.g. \cite{CastorLamers}).
Thus we adopted a generalized form of this law, $v\sim (1-1/x^{\alpha_1})^{\alpha_2}$.
In case of non-magnetic accretion disk winds, the azimuthal component of the velocity quickly becomes much smaller than the radial one, and such a  velocity profile describes the real distribution of the line-of-sight velocity reasonably well. 

During X-ray bursts, the situation can be more complicated.
During the burst, the temperature at the stellar photosphere may quickly approach $T\sim 10^9K$ and the electron scattering occurs in the Klein-Nishina regime. The reduced cross section allows for higher radiation flux (locally below Eddington) to diffuse out to larger radii where temperature is smaller and the radiation flux is locally supper Eddington. A quasi-stationary wind may result. Calculations of such winds must self-consistently account on the radiation transfer and relativistic corrections and demonstrate stellar type velocity profiles (\cite{QuinnPacz, Nobili94}.)

To approximate the explosively
accelerated plasma of X-ray bursters we adopt a linear velocity law. Making use of a Hubble law, ${\bf v}={\bf r}/t$ drastically simplifies the radiation transfer calculations in 
supernova shells (\cite{Karp77}). 

Two parameterized opacity laws and a linear, and
generalized form of the velocity law, $v\sim (1-1/x^{\alpha_1})^{\alpha_2}$
were considered.

Our results for all considered velocity and opacity distributions show that
in particular circumstances, i.e. proper velocity and opacity laws, strong gravity,
the observed line profile
consists of  two 
absorption troughs separated
by the emission component. The red-shifted absorption line can be weak and can be superimposed on the emission wing. Some of the strongest W-shape profiles were found for the linear velocity law.
Surprisingly enough, no W-shape profiles are found for a $v\sim\sqrt{1-1/x}$ law. In this case 
the red-shifted absorption almost exactly compensates for the blue-ward emission peak, producing a sharp edge in the emission
(Fig.~\ref{prof_stel1}).

However, in many other cases of different $\alpha_1$,$\alpha_2$,
W-shape profiles are also found 
(Fig.~\ref{prof_stel2} - \ref{prof_stel3}). 
If the opacity is distributed as $\tau\sim (1-v/V^\infty)^k$, ($k>0$), the red-shifted absorption line in most cases is superimposed on the red-shifted emission. However, if the opacity peaks at smaller velocities, $\tau\sim 1/v^k$, distinct W-shape profiles are found again. Thus, the results suggest that the most favorable conditions realize when the opacity peaks at large gravitational red-shifts and in such a case the velocity profile is probably less important.
The further the distribution of the opacity is from that the more important is the 
velocity law and the relative importance of the Doppler effect.

If some observed spectral feature is interpreted as being formed in the flow in a proximity of a strongly gravitating compact object, then our results suggest that the assumption of the distribution of the opacity (optical depth law) is the most critical one. Such distribution should be calculated simultaneously with the ionization balance calculations in the background of the pre-calculated hydrodynamical model.

The separation between red- and blue-shifted absorption
features is a function only of the dynamics and relative
importance of gravity in the line forming region. 

The most robust prediction of our model is the possibility 
of the double absorption trough as a result of a concurrence between gravitational red-shifting and Doppler effects. Our calculations suggest that 
the particular shape and intensity of the emission component which separates these absorption lines depend sensitively on assumed parameters.

Gravitationally red-shifted absorption lines form in places close to the compact object.  
The emission component is necessarily formed in plasma occupying much 
larger volume.  
Thus, these features are formed in places which are possibly strongly separated in space.
Thus, from the perspective of future observations, it would be interesting to look for correlated variability of different components of the profile between each other and with the continuum. 

\section*{Acknowledgments}
Most of this work have been made when the author was a postdoctoral fellow at the Max-Planck Institute for Nuclear Research (Heidelberg).
This research was supported in part by an appointment to the NASA Postdoctoral Program at the NASA Goddard Space Flight Center, administered by Oak Ridge Associated Universities through a contract with NASA.
I thank  G.S. Bisnovatyi-Kogan for the encouragement of this work and also Felix Aharonian and members of the High Energy Astrophysics Group
of the Max-Planck Institute for Nuclear Research for
discussions. 
I thank Tim Kallman for discussions and his suggestions and help regarding the style and structure of the  manuscript. 

%I would like to thank the referee for his/her many 
%constructive comments, which have lead to improvement of the manuscript.

\subsection{Subsection title}

\bsp

\label{lastpage}

\end{document}